\documentclass[a4paper,11pt]{article}
\pdfoutput=1 
\usepackage{jheppub} 

\usepackage[T1]{fontenc} 
\usepackage{appendix}
\usepackage{amsmath}
\usepackage{braket}
\usepackage{indentfirst}
\usepackage{comment}
\usepackage{tikz}
\usetikzlibrary{tikzmark}

\newlength{\mytextsize}

\newcommand{\JoinUp}[5]{\begin{tikzpicture}[remember picture,overlay,line width=0.025\mytextsize]
    \draw([shift={(#1\mytextsize,#2\mytextsize)}]pic cs:start#5) -- ++(0pt,0.7\mytextsize) -| ([shift={(#3\mytextsize,#4\mytextsize)}]pic cs:end#5);
    \end{tikzpicture}}

\title{\boldmath Modeling Compact Objects with EFT II: \\The Post-Newtonian Expansion}

\author[]{Irvin Martinez}

\affiliation[]{High Energy Physics, Cosmology \& Astrophysics Theory Group, Department of Mathematics \& Applied Mathematics, University of Cape Town, Cape Town, 7701, South Africa}

\emailAdd{mrtirv001@myuct.ac.za}

\abstract{Part 2 of 3 from the master's thesis: Modeling Compact Objects with Effective Field Theory. Using the Effective Field Theory framework for extended objects, we build the effective theory of a binary system made up of the most general compact objects in a theory of gravity as General Relativity with electrodynamics, objects which are described by their mass, spin, charge and their finite-size structure. We obtain the leading order post-Newtonian expansion to each of the relevant terms in the effective action that have been derived using the coset construction, where the covariant building blocks to build up the tower of invariant operators are derived from symmetry principles. Having matched the coefficients of the theory from the literature, we show the predictivity of our theory by obtaining well known post-Newtonian results on spinning extended objects, as well as on charged objects. Then, we bring new results on the polarizability and dissipation of charged spinning compact objects.}

\begin{document} 
\maketitle
\flushbottom

\section{Introduction}
\label{sec:intro}

The detection of gravitational waves (GWs) from compact object binaries \cite{Abbott:2016blz, TheLIGOScientific:2017qsa,LIGOScientific:2017ync,LIGOScientific:2017zic,LIGOScientific:2021qlt} are providing us the opportunity to test fundamental physics in the strong regime of gravity. With an increasing catalog of detected black holes (BHs) and neutron stars (NSs) \cite{LIGOScientific:2020kqk,LIGOScientific:2020ibl}, and sensitivity improvements of current \cite{2020LIGO} and future GW detectors \cite{Punturo:2010zz,Maggiore:2019uih,Barausse:2020rsu}, there is a potential to probe the internal structure of the compact objects, which can be achieved by matching the coefficients of the effective theory \cite{Martinez:2021mkl} from GW observations \cite{Flanagan:2007ix, Martinez:2020loq}.
 
Hence the need on improving the theoretical and numerical modeling of the dynamics of compact objects. The properties of their internal structure play an important role mostly in the late inspiral phase of the coalescence \cite{Martinez:2020loq}, before the plunge and merger (or collision) of the components of the binary. There are different approaches of modeling compact object interactions, and in particular during the inspiral, the dynamics are well described by the post-Newtonian (PN) approximation \cite{Blanchet:2006zz}, which reproduces the dynamics of the full theory of General Relativity to high accuracy until the last stable orbit (LSO) \cite{ Baker:2006ha,Boyle:2007ft}. In the PN expansion, one obtains general relativistic corrections to the dynamics as a perturbative series in terms of the expansion parameter, $v/c \ll 1$, with $v$ the relative velocity of the binary. For each $n$-PN order, the expansion of the Lagrangian is of order $v^{2n}$.

A very powerful framework that has led to the derivation of many new results on the dynamics of binary systems, is the Effective Field Theory (EFT) for extended objects in non-relativistic General Relativity \cite{Goldberger:2004jt,Goldberger:2005cd}.  In this worldline effective theory, extended objects are treated as point particles, with their properties and internal structure encoded in higher order invariant operators that are allowed by the symmetries of the objects. Within this framework, it is possible to build a tower of EFTs \cite{Goldberger:2006bd}, which follows a clear hierarchy of scales: $\ell \ll r \ll \lambda$ , with $\ell$, the radius of the compact object, $r$, the orbital separation of the binary, and $\lambda = r/v$, the wavelength of the GW radiation \cite{Goldberger:2004jt}. In this EFT approach, given the hierarchy of scales, we can organise the PN expansion in a systematic way by using the usual diagrammatic tools from EFTs. 

Since the introduction of the EFT for extended objects, in which finite size \cite{Goldberger:2004jt} and dissipative effects \cite{Goldberger:2005cd} are taken into account for non-spinning objects, the framework has gone through a remarkable progress in the past few years. For instance, it was shown in \cite{Kol:2007bc}, that the computations simplifies by considering the time dimension as compact, and parametrizing the gravitational interaction in terms of non-relativistic fields.  Furthermore, spinning \cite{Porto:2005ac,Levi:2015msa, Delacretaz:2014oxa}  and charged \cite{Patil:2020dme,Martinez:2021mkl} objects were introduced, and the effective theory that describes the long wavelength gravitational radiation from compact binary systems was derived \cite{Goldberger:2009qd}.\footnote{non-spinning BH electrodynamics were first considered in \cite{Goldberger:2004jt,Goldberger:2005cd}, in which charge is neglected, but the polarizability and dissipative effects are taken into account.} Tidal effects for spinning objects were considered in \cite{Porto:2005ac,Levi:2015msa, Endlich:2015mke}, while dissipative effects for slowly spinning in \cite{Porto:2007qi, Endlich:2015mke}, and for maximally spinning in \cite{Goldberger:2020fot}. Charged spinning compact objects were introduced in \cite{Martinez:2021mkl}, in which the polarizability and the electromagnetic dissipation of the charged spinning objects are considered as well. Nevertheless, in the latter, only a formulation of the effective theory containing such effects is made.    

Therefore, the purpose of this work is to obtain the leading order (LO) PN expansion of the theory of charged spinning compact objects derived from the coset construction  \cite{Martinez:2021mkl}, which has not been done elsewhere. In this effective theory, the covariant building blocks are derived by recognizing the symmetry breaking pattern that a compact object generates, from which we can build the invariant operators to build the effective action. Within this procedure, the coefficients of the derived effective action from a coset are to be matched from the known full theory, and ultimately from observations. In \cite{Martinez:2021mkl}, we have matched the coefficients from the known theories \cite{Goldberger:2004jt,Goldberger:2005cd,Porto:2005ac,Levi:2015msa, Patil:2020dme, Goldberger:2020fot}, which allows us to show the predictivity of our theory.

To connect to the results of the PN expansion for spinning extended objects \cite{Porto:2005ac,Levi:2015msa,Levi:2016ofk}, we introduce the spin degree of freedom as the conjugate of the relativistic angular velocity, and show that from construction, our effective theory contains all the necessary ingredients to obtain the well known LO results. After recovering such results on spinning objects, by including the invariant operators corresponding to the corrections due to the electrodynamics, we re-derive the 1 PN correction to the interaction of charged point particles in curved space-time \cite{Patil:2020dme} in terms of non-relativistic fields, which simplifies the computation. Then, by considering the finite-size structure of an extended object in an external electromagnetic field, we bring new LO PN results on the polarizability and dissipation of compact objects.

Although in our work we have focused mostly on the LO corrections of the different effects that play a role in the dynamics, the EFT framework for extended objects \cite{Goldberger:2004jt,Goldberger:2005cd} has been extensively used to obtain the state of the art results of the PN expansion. Before the introduction of the EFT for extended objects, the first PN correction dates back to Einstein \cite{EIH}, and extraordinary efforts led to the obtention of the 3 PN expansion to the dynamics without spin \cite{ Blanchet:2006zz}, but the computation of higher order terms became very challenging. 

Then, within the EFT framework, the known results up to 3 PN order were reproduced \cite{Foffa:2011ub}, and new results for the 4 PN \cite{Foffa:2011ub,Foffa:2019yfl} and 5 PN \cite{Foffa:2019hrb} corrections to the dynamics were derived. On spinning objects, the PN contributions from the spin-orbit and spin-spin coupling were derived \cite{Porto:2005ac}, which received further corrections in \cite{Levi:2008nh, Porto:2008jj, Porto:2008tb, Levi:2010zu, Porto:2010tr, Levi:2015msa}. The current state of the art of the PN expansion for spinning objects is the complete 4 PN corrections \cite{Levi:2016ofk}, and partial results up to 5 PN order \cite{Levi:2019kgk,Levi:2020lfn,Levi:2020kvb,Levi:2020uwu}. The PN expansion for charged objects was considered until recently \cite{Patil:2020dme}, in which partial results of the 1 PN correction were derived. A review on the PN expansion using EFT can be found in \cite{Porto:2016pyg, Levi:2018nxp}.

Our work should be seen as an application of the effective theory derived in \cite{Martinez:2021mkl}, and has the purpose of, altogether with \cite{Martinez:2021mkl}, building an intuitive understanding of the derivation of the relevant effects that play an important role in the dynamics of compact object interactions. By establishing the connection between the effective theories derived using the coset construction \cite{Delacretaz:2014oxa,Martinez:2021mkl}, to the ones currently used for obtaining state of the art PN results \cite{Goldberger:2004jt,Goldberger:2005cd,Porto:2005ac,Levi:2015msa,Levi:2016ofk, Patil:2020dme}, and showing its equivalence, we have elucidated on the foundations of not only the PN expansion, but in general of the effective theory that describes the allowed compact objects in the theory of General Relativity. 

Whether or not charged BHs in binary systems exists in nature remains an open question. Although in the literature has been argued that such objects will tend to be electrically neutral, it is well known that a BH with an accretion disk can generate a strong magnetic field \cite{EventHorizonTelescope:2019dse}. Moreover, NSs with strong magnetic fields have been observed with light \cite{Hewish:1968bj}, known as magnetars or pulsars.  Nevertheless, the theoretical implementation of electromagnetic charge into the currently used frameworks for gravitational waveform extraction for coalescing binaries has not been fully explored. Therefore, in the light of a high number of upcoming GW observations, our results on the finite-size structure of compact objects interacting electromagnetically, shed light onto accurate compact object modeling and probing what can be found in nature.

The outline of this paper is the following. In section \ref{sec:EFT}, we introduce the EFT and the building blocks derived in \cite{Martinez:2021mkl}, to then transform from the proper to the lab frame, in which the PN expansion is computed. In section \ref{sec:binary}, we build up the effective theory of a binary system following the hierarchy scales of an inspiraling binary made up of compact objects. In section \ref{sec:PN}, we parametrize our effective theory in a non-relativistic parametrization, and extract the relevant Feynman rules to obtain the dynamics. Then, we show the predictivity of our theory by computing the relevant Feynman diagrams to recover well known results, and then derive new results on the internal structure of charged spinning compact objects. Finally, in section \ref{sec:discussion}, we discuss our results and conclude.

Our conventions are the same as in \cite{Delacretaz:2014oxa, Martinez:2021mkl}, in which the indices, $\mu, \nu, \sigma, \rho ...$, denote spacetime indices, the indices, $a, b, c, d ...$, denote Lorentz indices, and the indices, $i, j, k, l ...$, denote the spatial components of Lorentz indices. Our metric signature is $\eta_{ab} = \mathrm{diag} (-1, +1, +1, +1)$. We work with units $c=1$, and when relevant, we include the $c$ factor at the end of the computation to stress the PN order at which effects enter into the dynamics.

\section{The Effective Theory of Compact Objects}
\label{sec:EFT}

Using the coset construction \cite{Coleman:1969,Ivanov:1981wn} in the framework of the EFT for extended objects \cite{Goldberger:2004jt, Goldberger:2005cd,Delacretaz:2014oxa}, in  \cite{Martinez:2021mkl} we have built the leading order, eletric-like parity, effective action for charged spinning compact objects, taking into account for the internal structure, such as the polarizability, tidal deformation and dissipation. The key idea in modeling compact objects as an EFT is that we can treat them as point particles, with the properties and internal structure encoded in higher order operators allowed by the symmetries of the compact object. 

In the proper frame, this is represented by the effective action

\begin{flalign}
\mathcal{S}_{CO} = \int \mathrm{d} \tau \left\{-mc^2 + \sum_n c_n  \tilde{\mathcal{O}}_n \right\},
\label{eq:effectiveaction}  
\end{flalign}

\noindent with $\tau$, the proper time, and where the first term describes a relativistic point particle with mass, $m$, and the second term, the sum over all possible higher order corrections. The coefficients, $c_n$, are the Wilson coefficients of the effective theory that are to be matched from the full known theory, and ultimately from GW observations. The invariant operators, $\tilde{\mathcal{O}}_{n}$, constitute the higher order corrections that are allowed by the symmetries, which take into account for the properties and internal structure of the extended object. These operators can be built from the covariant building blocks that are derived using the coset construction, by identifying the symmetry breaking pattern that such object generates \cite{Delacretaz:2014oxa, Martinez:2021mkl}. The sum is in principle infinite, and can be cut-off to the desired accuracy. 

The effective action in eq. (\ref{eq:effectiveaction}), is the one particle EFT in which the scale of the single compact object, determined by its radius, $\ell$, has been removed. All the UV physics are encoded in the coefficients, $c_n = c_n(\ell)$, while the operators, $\tilde{\mathcal{O}} = \tilde{\mathcal{O}} (r)$, will depend on the scale of the effective theory, $r$, the orbital scale of the binary. Therefore, the terms in the effective action scale as powers of, $\ell/r$, and in the non-relativistic limit, as powers of, $v/c$, with $v$, the relative velocity of the binary, from which we can obtain the PN order at which each of the terms enters into the dynamics. The coefficients used in this paper have been identified in \cite{Martinez:2021mkl} from \cite{Goldberger:2004jt,Goldberger:2005cd,Porto:2005ac,Levi:2015msa,Hinderer:2007mb,Patil:2020dme}.

\subsection{The Effective Action}

The effective action for charged spinning compact objects derived in \cite{Martinez:2021mkl}, has been constructed in the proper frame. For the purpose of computing the PN expansion and expanding over the velocity, $v$, the action in the lab frame is needed. Therefore, we start by pointing out the covariant building blocks and basic ingredients of the theory in the proper frame, and then transform to the lab frame. For the detailed derivation of the effective action we refer the reader to \cite{Martinez:2021mkl}, in which the covariant building blocks that are used to construct the tower of invariant operators are derived in detail.

\subsubsection*{The Proper Frame}

In the proper frame of the object, the effective action that describes the most general compact object allowed in a theory of gravity as General Relativity, which can be charged and spinning, is expressed as

\begin{flalign}
\begin{split}
\mathcal{S}_{eff} =  \mathcal{S}_{CO} + \mathcal{S}_{0},
\end{split}
\end{flalign}

\noindent with $S_{0}$, the interaction Einstein-Maxwell action, 

\begin{flalign}
\mathcal{S}_0 = \int \sqrt{-g} \; \mathrm{d}^4 x \left\{ -\frac{1}{4 \mu_0} F_{\mu \nu} F^{\mu \nu} +  \frac{1}{16 \pi G} R + .\;.\;.  \right\},
\label{eq:sinteraction}
\end{flalign}

\noindent with $F_{\mu \nu}  = \partial_{\mu} A^{}_{\nu} - \partial_{\nu} A^{}_{\mu}$, the electromagnetic field tensor, and $R$, the Ricci scalar. The worldline point particle action that describes the compact object, $S_{CO}$, reads \cite{Martinez:2021mkl}

\begin{flalign}
\begin{split}
\mathcal{S}_{CO} = \int \mathrm{d}\tau  &\left\{  -mc^2 + q \tilde{A}_{a}\tilde{u}^{a} + \frac{I}{4} \tilde{\Omega}_{ab} \tilde{\Omega}^{ab} +  I  \tilde{\Omega}_{ab} \tilde{a}^a \frac{\tilde{u}^b \tilde{u}_0}{\tilde{u}^2}  \right. \\
&\;\;+ n_{q,\Omega} \tilde{\Omega}_a \tilde{E}^a +  n_{g,\Omega} \tilde{\Omega}_a \tilde{\Omega}_b \tilde{E}^{ab} + n_{q} \tilde{E}_{a} \tilde{E}^{a}  \\
& \left. \;\; +  n_{g} \tilde{E}_{ab} \tilde{E}^{ab} +    \tilde{E}_{a}  \tilde{\mathcal{P}}^a +  \tilde{E}_{ab} \tilde{\mathcal{D}}^{ab} + \;  .\;.\;. \; \right\}, \\
\end{split}
\label{eq:actionproper}
\end{flalign}

\noindent  with, $\tilde{u}^a$, the velocity of the object, $\tilde{\Omega}_{ab}$, the relativistic angular velocity, and $\tilde{A}_a$, the gauge field of electromagnetism. The electric like parity tensors, $\tilde{E}^{}_{a}$ and $\tilde{E}^{}_{ab}$, correspond to the electromagnetic dipolar and gravitational quadrupolar moment respectively. The composite operators, $\tilde{\mathcal{P}}^a$ and $\tilde{\mathcal{D}}^{ab}$, encode the dissipative degrees of freedom. The action describing the compact object, lives in the worldline, while the underlying theory, eq. (\ref{eq:sinteraction}), lives in the bulk. 

The first three terms of eq. (\ref{eq:actionproper}) describes the LO correction of a charged spinning point particle, with mass, $m$, charge, $q$, moment of inertia $I$, and angular velocity, $\tilde{\Omega}^{ab}$. The fourth term is a higher order correction to the spin which is necessary for a correct description of the dynamics at the order that we consider. The first term in the second line is a spin correction due to the electromagnetic coupling, while the second term is a spin correction due to gravity. The last term of the second line corresponds to the coupling between electromagnetism and the ensuing dipole, while the first term in the last line corresponds to the coupling between gravity and the ensuing quadrupole. The last two terms couple the degrees of freedom which are responsible for the electromagnetic and gravitational dissipation during an interaction.

All the terms from the second and third line, encode the fact that a compact object is not truly a point particle, and therefore are finite-size corrections \cite{Goldberger:2004jt, Goldberger:2005cd}. We refer to the size effects due to gravity as tidal effects, while for the ones due to electromagnetism, we refer as polarization effects. The internal structure of the compact object is encoded in the coefficients of the theory, $n_{q},\, n_{g},\,n_{q,\Omega}$ and $n_{g,\Omega}$, and the coefficients, $c_q$ and $c_{g}$, which come from the operators that encode the dissipative effects. The explicit expressions of the operators due to dissipative effects with their coefficients, which we show below, can be obtained using the in-in formalism \cite{Jordan:1986ug}. These coefficients, as well as the ones from the corrections from the first line, $n_{\Omega} = I/4$ and $n_{\Omega,a} = I$, have been identified from the literature in \cite{Martinez:2021mkl}. 

The explicit expressions of the covariant building blocks included in the effective action reads \cite{Martinez:2021mkl}

\begin{flalign}
\tilde{u}^a & = \partial_{\tau} x^a, \\
\tilde{A}^a & = \Lambda_{b}^{\;\;a} A^b = \Lambda_{b}^{\;\;a}  e_{\mu}^{\;b} A^{\mu}, \\
\tilde{a}^a &= \partial_{\tau} \tilde{u}^a + \tilde{u}^{\mu} \omega_{\mu\;\;}^{\;\;ab} \tilde{u}_b \\
\tilde{\Omega}^{ab} & =  \Lambda_{c}^{\;\;a} (\eta^{cd} \partial_{\tau} + \tilde{u}^{\mu} \omega_{\mu}^{\; \; cd})\Lambda_{d}^{\;\;b}, \label{eq:spinp}\\
\tilde{E}^a & = \tilde{F}^{ab} \tilde{u}_b,\\
\tilde{E}^{ab} & = \tilde{W}^{acbd} \tilde{u}_c \tilde{u}_d, \\
\tilde{P}^{a} & = ic_{q} \dot{\tilde{E}}^{a},  \label{eq:dissep} \\ 
\tilde{\mathcal{D}}^{ab} & = ic_{g}  \dot{\tilde{E}}^{ab},  \label{eq:dissgp} 
\end{flalign}

\noindent with $A_{\mu}$, being invariant under the gauge transformation, $A_{\mu} \rightarrow A_{\mu}  + \partial_{\mu}\xi$, and $\tilde{\Omega}^{0i} = \tilde{a}^0 = 0$, by definition. The relativistic angular velocity can be expressed in terms of the epsilon tensor, $\tilde{\Omega}_{ab} = \epsilon_{abc} \tilde{\Omega}^c$. The Lorentz matrices, $\Lambda^{\;\;b}_{a} = B^{\;\;c}_{a} (\beta^i) \mathcal{R}_{c}^{\;\;b }(\theta^i)$, are parametrized by the product of a boost and a rotation matrix, with $\beta^i$, the velocity as a function of the rapidity, and $\theta^i$, the Euler angles that describe the orientation of the compact object \cite{Delacretaz:2014oxa,Martinez:2021mkl}. 

The field, $e^{a}_{\mu}$, is the vierbein from the tetrad formalism, which defines the metric as, $g_{\mu \nu} = \eta_{ab} e_{\mu}^a e_{\nu}^b$. The vierbein is used as well to change from the local to the general orthogonal frame, i.e. $A_{\mu} = e_{\mu}^{\; b} A_{b}$, and vice versa. The spin connection, $\omega_{\mu}^{ab}$, is defined in terms of the vierbein field \cite{Delacretaz:2014oxa},

\begin{equation}
\omega_{\mu}^{ab} (e) = \frac{1}{2} \left\{ e^{\nu a} (\partial_{\mu} e_{\nu}^{\;\;b} - \partial_{\nu} e_{\mu}^{\;\;b}) + e_{\mu c} e^{\nu a} e^{\lambda b} \partial_{\lambda} e_{\nu}^{\;\;c} - (a \leftrightarrow b)\right\}.
\label{eq:spinc}
\end{equation}

\noindent The Christoffel symbol in terms of the vierbein and the spin connection is identified \cite{Martinez:2021mkl},

\begin{flalign}
\Gamma_{\mu \nu}^a = \partial_{\mu} e^{\;a}_{\nu} + e_{\mu b} \omega_{\nu}^{ab},
\label{eq:cristoffele}
\end{flalign}

\noindent from which one can recover the space-time Christoffel symbol, $ \Gamma^{\sigma}_{\mu \nu} = e_{\; a}^{\sigma} \Gamma_{\mu \nu}^a $, or obtain the spin connection in terms of the Christoffel symbol.

The electric parity tensors in the proper frame,  $\tilde{E}^{}_{a}$ and $\tilde{E}^{}_{ab} $, for the electromagnetic and gravitational case, depend on the electromagnetic strength field tensor $\tilde{F}_{ab}$, and on the Weyl tensor, $\tilde{W}_{abcd}$, respectively. The dissipative operators in eqs. (\ref{eq:dissep}) and (\ref{eq:dissgp}), are the in-in expectation values at the initial state of the internal degrees of freedom, which are defined through the in-in path integral \cite{Jordan:1986ug}, and which are in general a function of the building blocks, $\tilde{E}^{a}$ and  $\tilde{E}^{ab}$, respectively  \cite{Goldberger:2005cd,Goldberger:2020fot}. The dot denotes derivative with respect to $\tau$. The dissipative degrees of freedom are encoded in the imaginary part of the action.

\subsubsection*{The Lab Frame}

In our effective theory, the position of the particle, $x^{\mu}$, is not affected by the local Poincaré group, but it is transformed under diffeomorphisms \cite{Delacretaz:2014oxa}. To transform the rest of the building blocks from the proper frame to the lab frame, we use the Lorentz matrices, $\Lambda^{a}_{\;\;b} = B^{a}_{\;\;c} (\beta^i) \mathcal{R}_{\;\;b}^{c }(\theta^i)$. The covariant building blocks in the lab frame are given by 

\begin{flalign}
v^a &= \partial_{t} x^a, \\
A^a & = \Lambda^{a}_{\;\; b} \tilde{A}^b  ,  \\
a^a &= \Lambda_{\;\;b}^{a} \tilde{a}^{b} = (\partial_{\tau} v^a + v^{\mu} \omega_{\mu}^{\;\;ab} v_b) \\
\Omega^{ab} & = \Lambda^{a}_{\;\;c} \Lambda^{b}_{\;\;d} \tilde{\Omega}^{cd}  =   \Lambda^{\;\; a}_{c} \partial_{t} \Lambda^{cb} + v^{\mu} \omega_{\mu}^{\;\;ab}, \label{eq:angularbb}  \\
E^a & = \Lambda^{a}_{\;\; b} \tilde{E}^b   = F^{ac}v_c,\\
E^{ab} & = \Lambda^{a}_{\;\;c} \Lambda^{b}_{\;\;d} \tilde{E}^{cd}  = W^{acbd} v_c v_d,\\
\mathcal{P}^a &=\Lambda^{a}_{\;\; b}\tilde{\mathcal{P}}^{b}, \\
\mathcal{D}^{ab} &= \Lambda^{a}_{\;\;c} \Lambda^{b}_{\;\;d} \tilde{\mathcal{D}}^{cd},\\
\label{eq:buildingblockslab2}    
\end{flalign}

\noindent  with $t$, the time measured by the observer, and the dot denotes derivative with respect to $t$. Given the usual notation of the PN expansion, we have denoted the velocity,  $u^a = v^a$, with $v^0 =1$. Having transformed the covariant quantities, the worldline action that describes the compact object in the lab frame now reads,

\begin{flalign}
\begin{split}
\mathcal{S}_{CO} = \int \mathrm{d}t  &\left\{  -m + q A_{a} v^{a} + \frac{I}{4} \Omega_{ab} \Omega^{ab} +  I\Omega_{ab} \frac{a^a v^b }{v^2}  \right.\\ 
& \;\;\,+ n_{q,\Omega} \Omega_a E^{a} +  n_{g,\Omega} \Omega_a \Omega_b E^{ab}  +n_{q} E_{a} E^{a}\\
&\left. \;\;\,  + n_{g} E_{ab} E^{ab}  +  E_{a}\Lambda^{a}_{\;\; b} \tilde{\mathcal{P}}^b   + E_{ab} \Lambda^{a}_{\;\;c} \Lambda^{b}_{\;\;d} \tilde{\mathcal{D}}^{cd}  \, +   .\;\;.\;\;.  \right\}. \\
\label{eq:actionlab}
\end{split}
\end{flalign}

To obtain the PN expansion for spinning objects, it is necessary to introduce the relativistic spin degree of freedom \cite{Porto:2005ac,Levi:2015msa,Steinhoff:2021dsn},

\begin{flalign}
S^{ab} = 2 \frac{\partial \mathcal{L}}{\partial \Omega_{ab}},
\end{flalign}

\noindent which is the conjugate variable of the angular velocity. Then, by Legendre transforming the Lagrangian that describes the compact object, we obtain

\begin{flalign}
\mathcal{L}_{CO} = -m + \frac{1}{2} S_{ab} \Omega^{ab} + .\;.\;. = -m + \frac{I}{2} S_{ab} ( \Lambda^{a}_{\;\;c} \partial_{\tau} \Lambda^{cb} +  \partial_{t} x^{\mu} \omega_{\mu}^{\;ab}),  
\label{eq:lomega2}
\end{flalign}

\noindent where we have inserted the explicit expression of the angular velocity, eq. (\ref{eq:angularbb}). This is the minimal spin coupling, and the ellipses denote the rest of the terms, which are taken into account when obtaining the PN expansion. The spin minimal coupling correction, eq. (\ref{eq:lomega2}),
 is equivalent to the one in \cite{Levi:2015msa,Porto:2016pyg}.


\section{Binary Inspiral}
\label{sec:binary}

The effective action, eq. (\ref{eq:actionlab}), describes the EFT of a single compact object, with their coefficients, $n(\ell)$, encoding the internal structure. For a compact object, $\ell \sim G m$. To describe a binary
system, we construct the effective action

\begin{equation}
    \mathcal{S}_{eff} = \mathcal{S}_0 + \sum_{n = 1}^2 \mathcal{S}_{CO},
    \label{eq:fullactionco}
\end{equation}

\noindent where the first term, $\mathcal{S}_{0} = \mathcal{S}_g + \mathcal{S}_q$, is the Einstein-Maxwell action in eq. (\ref{eq:sinteraction}), and the second term is the action describing each of the compact objects that form the binary system, with $\mathcal{S}_{CO}$ the worldline point particle action, eq. (\ref{eq:actionlab}). The operators that appear in $\mathcal{S}_{CO}$, depend only on the scale of the binary, $r$, the orbital separation between the two compact objects. At the scale of the inspiraling binary, which is bound gravitationally, the virial theorem holds, $Gm/r \sim v^2$ \cite{Goldberger:2004jt}. Therefore, we encounter the first hierarchy of scales, $\ell \ll r$.

The gravitationally bound system composed of compact objects, will shrink its orbit due to the loss of energy and momentum through gravitational wave radiation, whose frequency is fixed by the orbital frequency of the binary, $\omega_s$. The emitted radiation is ultimately observed by our GW detectors, which implies that these radiation modes are on shell, satisfying the momentum relation, $k_0 = \vec{k}$, for its four momentum vector, $k = (k_0,\vec{k}) $. This is in contrast to the modes that mediate the gravitational interaction, which are off-shell and therefore its momentum, $k_0 \neq \vec{k}$. Therefore, we can establish the next hierarchy, $r \ll \lambda$, with $\lambda$ the wavelength of radiation, scaling as $\lambda \sim 1/\omega_s \sim r/v$. Thus, an inspiraling binary made up of compact objects, satisfies the  hierarchy of scales  \cite{Goldberger:2004jt},

\begin{flalign}
	\ell \ll r \ll \lambda.
\end{flalign} 

Given this hierarchy, we expand the gravitational field around flat spacetime \cite{Goldberger:2004jt}, 

\begin{flalign}
g_{\mu \nu} = \eta_{\mu \nu} + H_{\mu \nu} + h_{\mu \nu},
\end{flalign}

\noindent where $H_{\mu \nu }$, are the potential modes that mediate the interaction between the two objects, and $h_{\mu \nu}$, the radiation modes, which are finally observed by the GW detectors. The latter modes are on-shell, while the former ones are off-shell, which imply that its components scale as \cite{Goldberger:2004jt}

\begin{flalign}
\partial_{t} H_{\mu \nu} \sim \frac{v}{r}H_{\mu \nu}, \;\;\; &\;\; \; \partial_i H_{\mu \nu} \sim \frac{1}{r} H_{\mu \nu}, \label{eq:scalingH}\\
h_{\mu \nu} \sim& \frac{v}{r} h_{\mu \nu}.
\label{eq:scalingh}    
\end{flalign}

\noindent To obtain the conservative dynamics of the binary system, we are interested in integrating out the orbital modes. 

The photon field, $A_{\mu}$, is decomposed analogously to the gravitational field. It is separated into \cite{Patil:2020dme}

\begin{flalign}
A_{\mu} = \textbf{A}_{\mu} + \bar{A}_{\mu},
\end{flalign}

\noindent with $\bar{A}_{\mu}$, the radiation modes, and $\mathbf{A}_{\mu}$, the potential modes that mediate the interaction. Their scaling is equivalent to the ones in eq.  (\ref{eq:scalingH}) and (\ref{eq:scalingh}), for each mode respectively, 

\begin{flalign}
\partial_{t} \mathbf{A}_{\mu} \sim \frac{v}{r}\mathbf{A}_{\mu}, \; \; \; & \;\; \; \partial_i \mathbf{A}_{\mu} \sim \frac{1}{r} \mathbf{A}_{\mu},\label{eq:scalingelectroA}\\ 
\bar{A}_{\mu} \sim & \frac{v}{r} \bar{A}_{\mu}.
\label{eq:scalingelectroa}    
\end{flalign}

Therefore, the conservative dynamics of the interaction of a binary composed of charged spinning compact objects, is obtained in a diagrammatic approach by integrating out the potential modes of the gravitational and electromagnetic field, which removes the orbital scale of the binary. The procedure of integrating out the orbital scale, is defined by the functional integral

\begin{flalign}
	e^{i\mathcal{S}_{bin}} = \int D H_{\mu \nu}  D \mathbf{A}_{\mu} e^{i \mathcal{S}_{eff}},
\end{flalign}

\noindent with, $\mathcal{S}_{eff}$, the action defined in eq. (\ref{eq:fullactionco}). The action, $\mathcal{S}_{bin}$, describes the conservative dynamics of the binary system. In this interaction, only tree diagrams contribute to the dynamics, given that these objects are described classically, and that any quantum effects \cite{Donoghue:1994dn, Bjerrum-Bohr:2002gqz}, at least during the inspiral, can be neglected. After integrating out the orbital modes, the effective theory of the binary system will be given by the action,

\begin{equation}
    \mathcal{S}_{bin} = \mathcal{S}_0 (\eta_{\mu \nu} + h_{\mu \nu},\bar{A}_{\mu}) + \sum_{n = 1}^2 \mathcal{S}_{CO} (\eta_{\mu \nu} + h_{\mu \nu},\bar{A}_{\mu}),
    \label{eq:actioncointegratedout}
\end{equation}

\noindent which is often expressed as one composite body, 

\begin{equation}
    \mathcal{S}_{bin} = \mathcal{S}_0 (\eta_{\mu \nu} + h_{\mu \nu},\bar{A}_{\mu}) +  \mathcal{S}_{OB} (\eta_{\mu \nu} + h_{\mu \nu},\bar{A}_{\mu}).
    \label{eq:actionbin}
\end{equation}

\section{The Post-Newtonian Expansion}
\label{sec:PN}

In this work, rather than to obtain high order PN corrections, we show the predictivity of our theory by recovering well known results, and then derive new ones on the finite-size structure of charged spinning compact objects. The computation of high order PN corrections for such objects will be approached in an upcoming work. In this section we will be as explicit as possible on the derivation of the Feynman rules and the computation of the Feynman diagrams, to provide a review on the basic description of a binary system composed of compact objects. For a thorough review on the PN expansion, including higher order PN corrections, we refer the reader to \cite{Porto:2016pyg,Levi:2018nxp}. The code EFTofPNG package for computing the post-Newtonian expansion \cite{Levi:2017kzq}, was extensively used to cross check results.

\subsection{Non-Relativistic General Relativity}

In the non-relativistic regime of gravity \cite{Goldberger:2004jt},  the time dimension can be regarded as compact in comparison to the spatial dimensions \cite{Kol:2007bc}. Thus, we switch to a Kaluza-Klein (KK) parametrization, with the metric \cite{Kol:2007bc},

\begin{equation}
    g_{\mu \nu} \mathrm{d}x^{\mu} \mathrm{d}x^{\nu} \equiv - e^{2\phi} (\mathrm{d}t - \mathcal{A}_i \mathrm{d}x^i)^2 + e^{-2\phi} \gamma_{ij}\mathrm{d}x^i \mathrm{d}x^j,
\end{equation}

\noindent which defines the KK or non-relativistic fields: $\phi$, the Newtonian scalar, $\mathcal{A}_i$, the gravitomagnetic vector of the KK parametrization, and $\gamma_{ij} \equiv \delta_{ij} + \sigma_{ij}$, the tensor field with, $\gamma^{ij} \gamma_{jk} \equiv \delta^{i}_{k}$ and $\mathcal{A}^i \equiv \gamma^{ij} \mathcal{A}_j$. The matrix form of the metric is given by,

\begin{equation}
    g_{\mu \nu} = 
    \begin{pmatrix}
    -e^{2 \phi} & e^{2\phi} \mathcal{A}_j \\ 
    e^{2\phi} \mathcal{A}_{i} \; \; \; & e^{-2 \phi} \gamma_{ij} - e^{2 \phi} \mathcal{A}_{i} \mathcal{A}_{j} 
    \end{pmatrix}.
    \label{eq:metricKK}
\end{equation} 

From eq. (\ref{eq:metricKK}), the vierbein in the KK parametrization can be read off. As pointed out in \cite{Levi:2015msa}, the vierbein can be gauge fixed using the time gauge of Schwinger \cite{Schwinger:1963re}, which fixes the time axes of the local coordinate system to the one of the general coordinate frame. In equations, we fix $e^{\;a}_0$, and choose a local frame in which their spatial components are zero,  

\begin{flalign}
e_{0}^{i} = 0. 
\end{flalign}

\noindent Using this constraint, altogether with eq. (\ref{eq:metricKK}), and $g_{\mu \nu} = \eta_{ab} e_{\mu}^a e_{\nu}^b$, the vierbein reads \cite{Levi:2015msa}

\begin{equation}
    e_{\mu}^{a} =  
    \begin{pmatrix}
    e^{\phi} & -e^{\phi} \mathcal{A}_i \\ 
    0 & e^{- \phi} \sqrt{\gamma}_{ij} 
    \end{pmatrix},
\end{equation} 

\noindent which completely fixes the redundant degrees of freedom.

To extract the Feynman rules and obtain the propagators of the theory, we need to fix the gauge of the interaction action, $\mathcal{S}_0$, to remove the physically equivalent field configurations. For the gravitational action, $\mathcal{S}_{g}$, we choose the harmonic gauge \cite{Goldberger:2004jt}

\begin{equation}
    \mathcal{S}_{g} = \mathcal{S}_{EH} + \mathcal{S}_{GF,g} =  \frac{1}{16 \pi} \int \mathrm{d}^4 x \sqrt{-g} R -  \frac{1}{32 \pi G} \int \mathrm{d}^4x\sqrt{-g}g_{\mu \nu} \Gamma^{\mu} \Gamma^{\nu},
\end{equation}

\noindent with $\Gamma^{\mu} \equiv \Gamma^{\mu}_{\rho \sigma} g^{\rho \sigma}$. Then, by parametrizing $\mathcal{S}_g$ in a KK fashion, the gravitational action can be cast as \cite{Kol:2007bc}

\begin{flalign}
\mathcal{S}_g = \frac{1}{16 \pi G} \int \mathrm{d} t \mathrm{d}^3  x \sqrt{\gamma} \left( R[\gamma] - 2 \partial_{i} \phi \partial^{i} \phi + \frac{1}{4} e^{4 \phi} \mathcal{F}^2 \right),
\label{eq:KKgravity}
\end{flalign}

\noindent  with $\partial_{i} \partial^i \phi = \gamma^{ij} \partial_i \phi \partial_j \phi$, and where we have defined the analog of the field strength tensor, $F_{ab}$, but in terms of the KK gravitomagnetic tensor, $\mathcal{A}_i$, such that, $\mathcal{F}_{ij} = \partial_i \mathcal{A}_j - \partial_j \mathcal{A}_i$. On the action for the photon field,

\begin{flalign}
	\mathcal{S}_q = \mathcal{S}_{M} + \mathcal{S}_{GF,q} = - \frac{1}{\mu_0} \int \mathrm{d}^4 x \sqrt{-g} \left( \frac{1}{4} F^2  + \frac{1}{2 \xi}(\partial_{\mu} A^\mu)^2 \right),
	\label{eq:selectrogf}
\end{flalign}

\noindent we choose the Feynman-'tHooft gauge, which  implies $\xi = 1$. 

\subsection{The Feynman Rules}

We start by extracting the propagators of the effective theory. From the gravitational action in the KK decomposition, eq. (\ref{eq:KKgravity}), the propagators can be read off from the terms that are quadratic in the fields. The non-relativistic field propagators in the harmonic gauge reads \cite{Kol:2007bc} 

\begin{align}
    & \begin{gathered}
     \includegraphics[width =2cm]{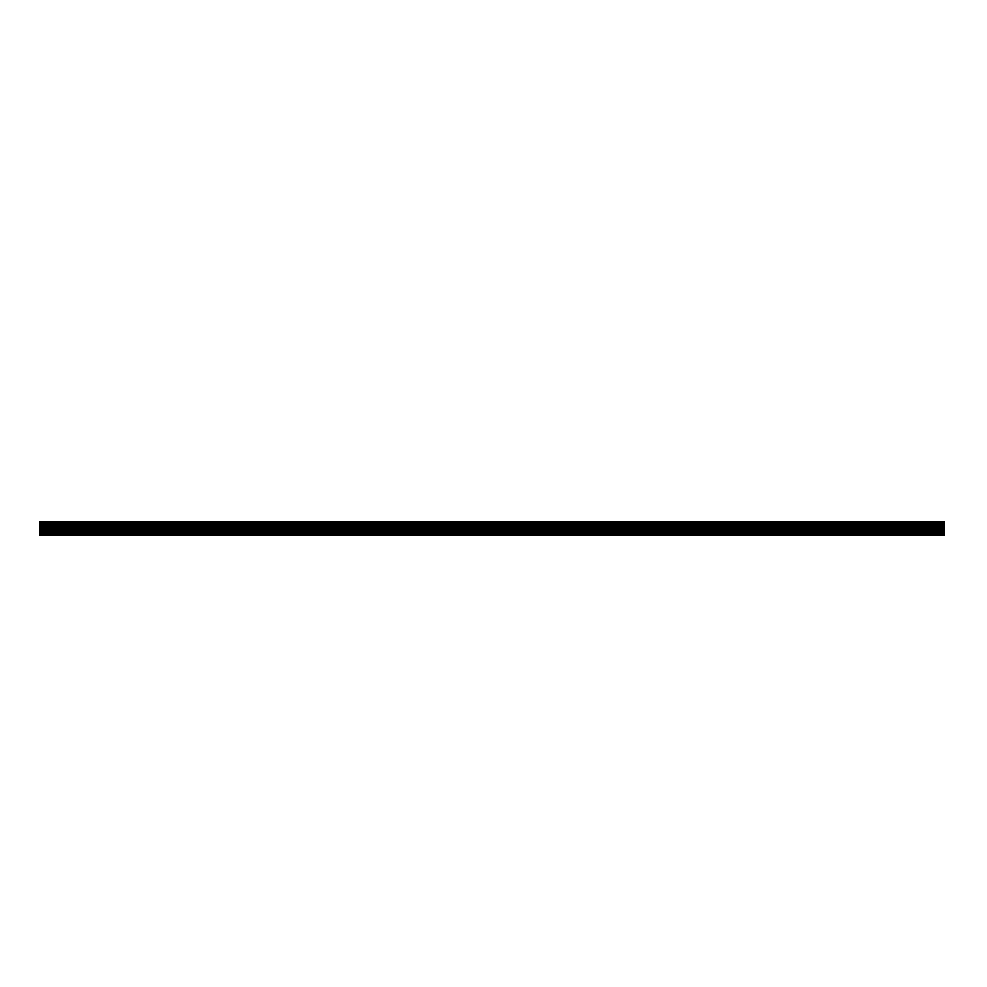}
    \end{gathered} = \left< \phi (x_1) \phi (x_2)\right> = 4 \pi G \; \delta(t_1 - t_2)  \int \frac{\mathrm{d}^3 \vec{k}}{(2\pi)^3} \frac{e^{i \vec{k}\cdot \vec{r}}}{\vec{k}^2} , \label{eq:propN}\\
    & \begin{gathered}
     \includegraphics[width =2cm]{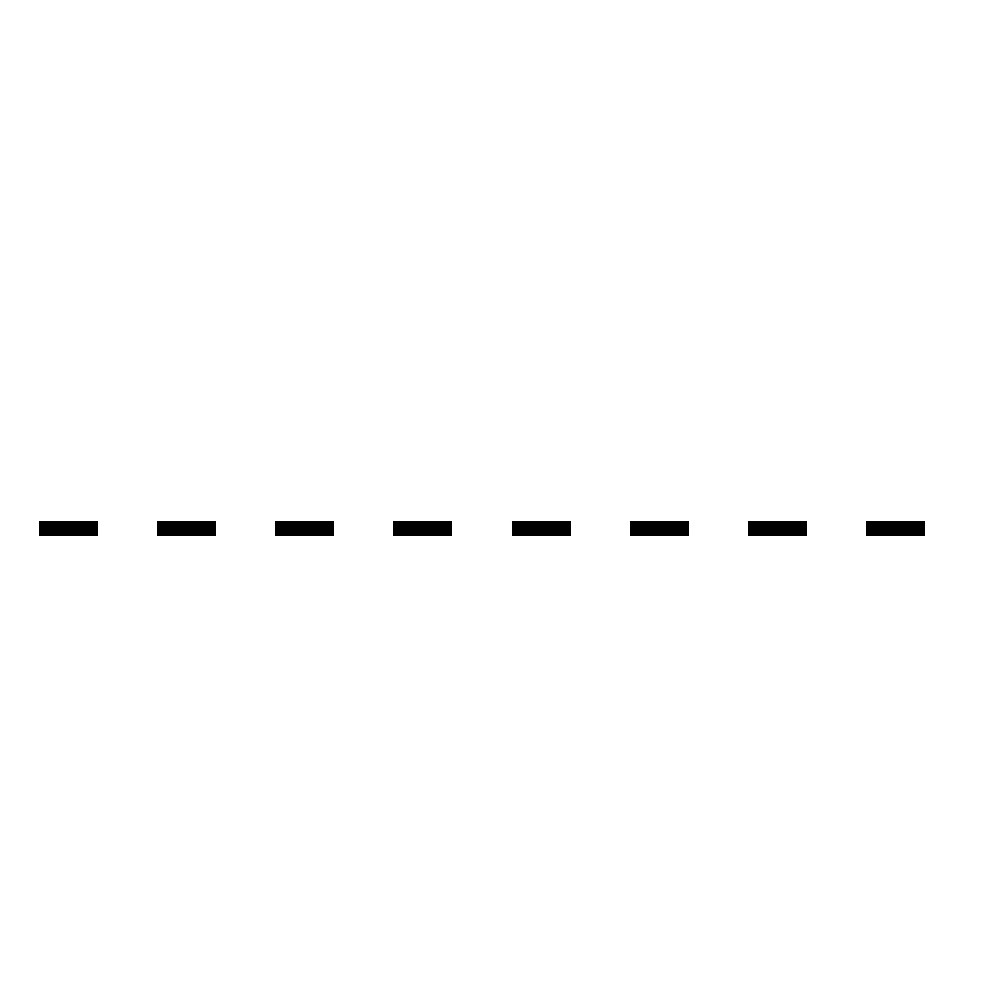}
    \end{gathered} = \left< \mathcal{A}_i (x_1) \mathcal{A}_j(x_2)\right> = - 16\pi G \; \delta (t_1 - t_2)  \int \frac{\mathrm{d}^3 \vec{k}}{(2\pi)^3} \frac{e^{i \vec{k}\cdot \vec{r}}}{\vec{k}^2} \delta_{ij},\\
    & \begin{gathered}
     \includegraphics[width =2cm]{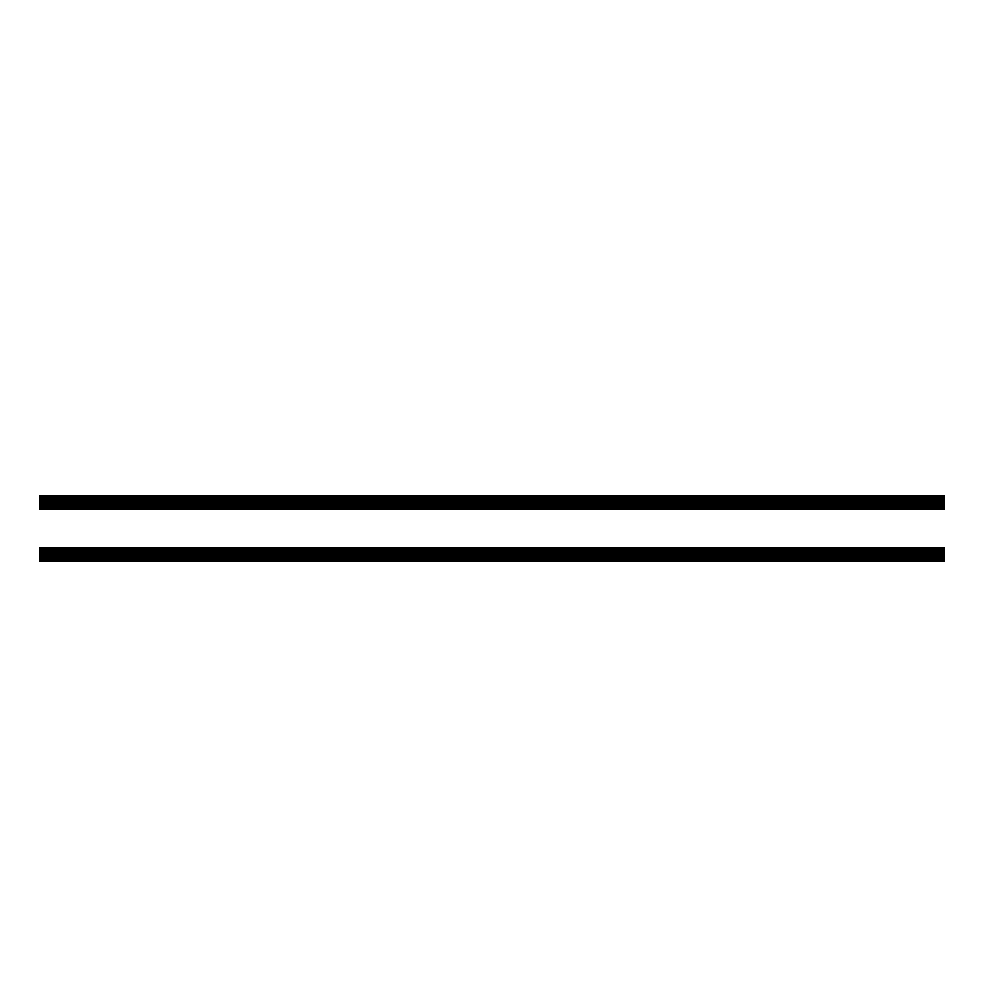}
    \end{gathered} = \left< \sigma_{ij} (x_1) \sigma_{kl}(x_2)\right> =  32 \pi G \; \delta (t_1 - t_2)  \int \frac{\mathrm{d}^3 \vec{k}}{(2\pi)^3} \frac{e^{i \vec{k}\cdot \vec{r}}}{\vec{k}^2} P_{ij;kl},
\end{align}

\noindent with $P_{ij;kl} = \frac{1}{2} (\delta_{ik} \delta_{jl} + \delta_{il} \delta_{jk} - 2 \delta_{ij} \delta_{kl})$.  We have used the scaling of the orbital modes, $k_0 \simeq v/r$ and $|\vec{k}|\simeq 1/r$, and expanded over, $k^2_0/\vec{k}^2 \sim v^2$, such that

\begin{flalign}
\int \frac{\mathrm{d}^4 k}{(2\pi)^4} e^{-i \vec{k} \cdot \vec{r}}\frac{1}{k^2} = \int \frac{\mathrm{d}k_0}{2 \pi} \frac{\mathrm{d}^3 \vec{k}}{(2 \pi)^3} e^{-ik_0} \frac{e^{i \vec{k} \cdot \vec{r}}}{\vec{k}^2} \left( 1 + \frac{k^2_0}{\vec{k}^2} + .\;.\;. \right) = \delta(t) \int \frac{\mathrm{d}^3k}{(2\pi)^3}  \frac{e^{i\vec{k}\cdot \vec{r}}}{\vec{k}^2},
\label{eq:kexp}
\end{flalign}

\noindent and therefore the propagator is instantaneous, with the relativistic time corrections to the propagator being considered as quadratic perturbations \cite{Goldberger:2004jt}. The propagators obtained in the KK parametrization have a simple form, in which the fields, $\phi$ and $\mathcal{A}_i$, dominates the interaction. The time relativistic corrections can be read from eq. (\ref{eq:propN}) and (\ref{eq:kexp}),

\begin{align}
    & \begin{gathered}
     \includegraphics[width =2cm]{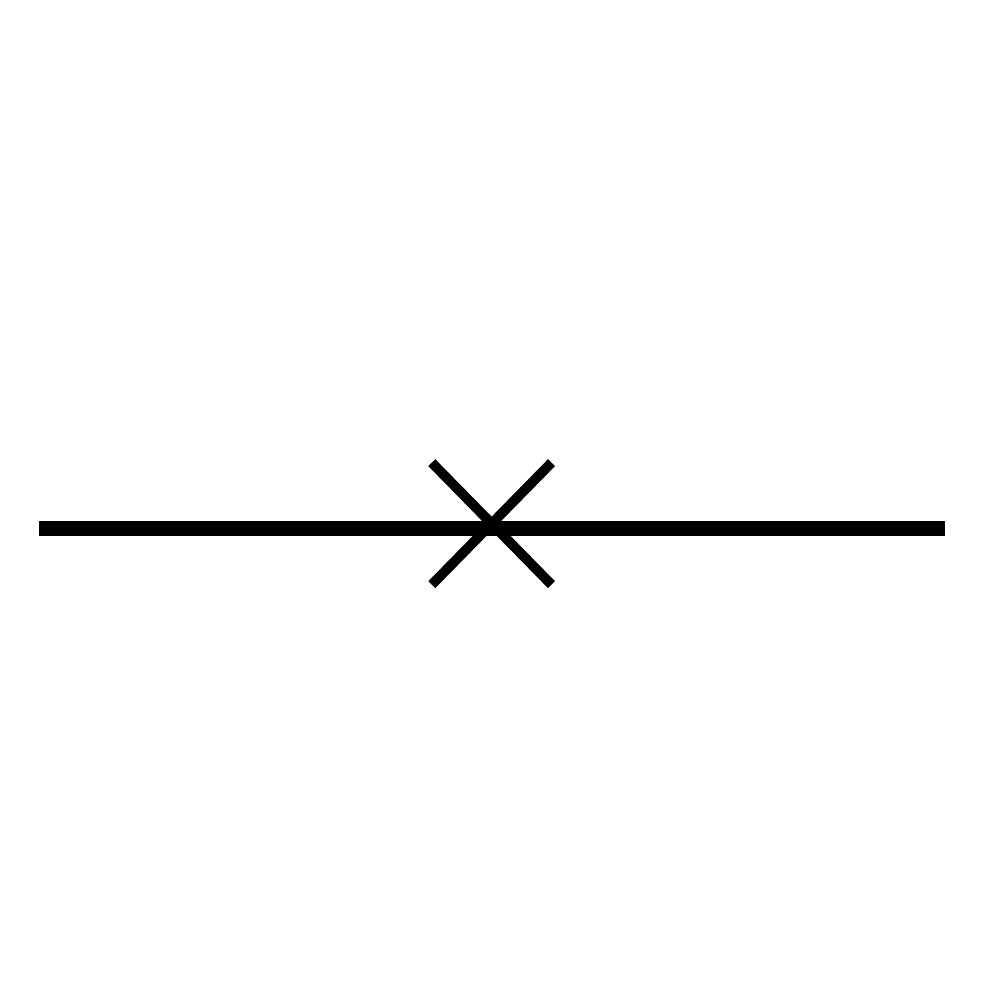}
    \end{gathered}  =  4 \pi G \;  \frac{\mathrm{d}}{\mathrm{d}t_1 \mathrm{d}t_2} \delta(t_1 - t_2) \int \frac{\mathrm{d}^3 \vec{k}}{(2\pi)^3} \frac{e^{i \vec{k}\cdot \vec{r}}}{\vec{k}^4}.
\end{align}

From the electromagnetic interaction action in a non-relativistic parametrization, the propagators can be read off from the quadratic terms as well. We separate the time dimension of the photon field from the spatial ones, such that the propagators reads

\begin{align}
    & \begin{gathered}
     \includegraphics[width =2cm]{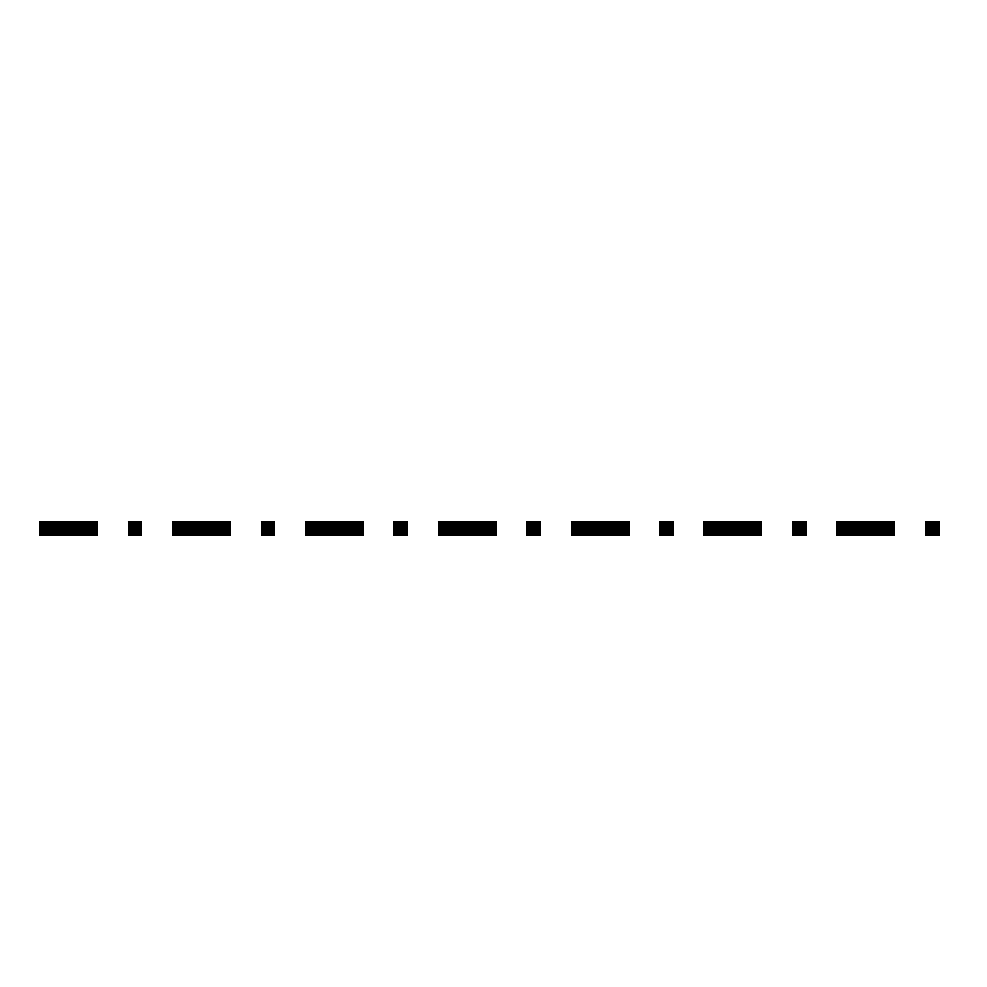}
    \end{gathered} = \left< A_0 (x_1) A_0 (x_2)\right> = -  \mu_0 \; \delta(t_1 - t_2)  \int \frac{\mathrm{d}^3 \vec{k}}{(2\pi)^3} \frac{e^{i \vec{k}\cdot \vec{r}}}{\vec{k}^2} ,\\
    & \begin{gathered}
     \includegraphics[width =2cm]{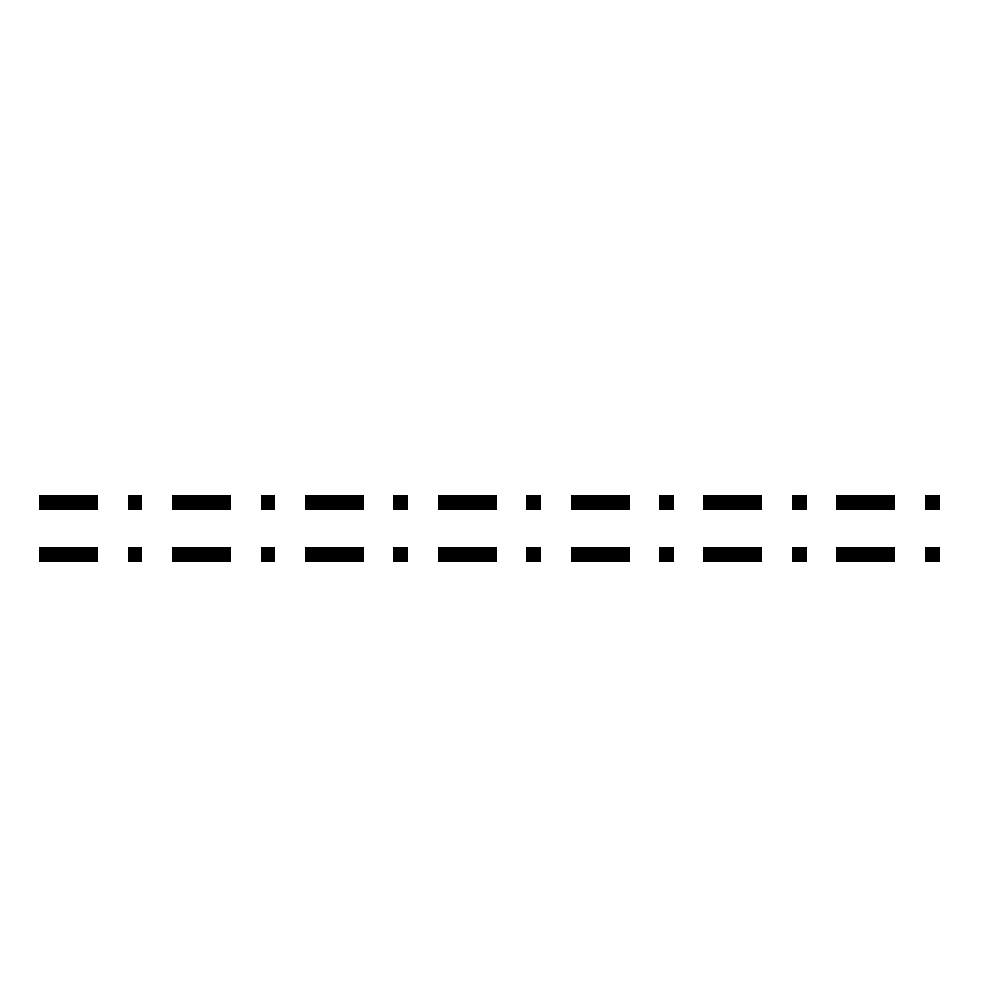}
    \end{gathered} = \left< A_i (x_1) A_j(x_2)\right> =  \mu_0 \; \delta (t_1 - t_2)  \int \frac{\mathrm{d}^3 \vec{k}}{(2\pi)^3} \frac{e^{i \vec{k}\cdot \vec{r}}}{\vec{k}^2} \delta_{ij}.
\end{align}

\noindent The time relativistic correction,

\begin{align}
    & \begin{gathered}
     \includegraphics[width =2cm]{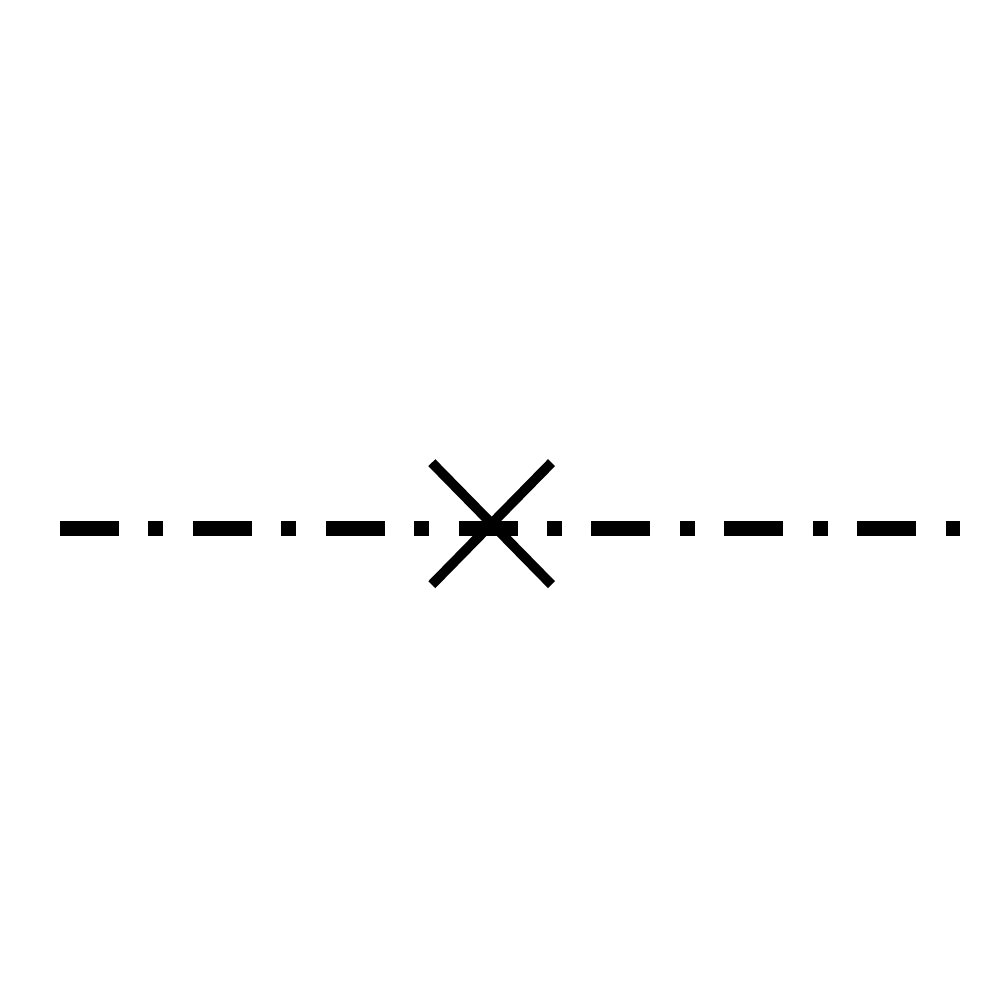}
    \end{gathered}  =  -  \mu_0 \; \frac{\mathrm{d}}{\mathrm{d}t_1 \mathrm{d}t_2} \delta(t_1 - t_2) \int \frac{\mathrm{d}^3 \vec{k}}{(2\pi)^3} \frac{e^{i \vec{k}\cdot \vec{r}}}{\vec{k}^4}.
\end{align}

Having derived the relevant propagators, we express the worldline action of the binary system in terms of the non-relativistic fields, from which we can read off the vertices. We do so by considering term by term from the action in eq. (\ref{eq:actionlab}).

\subsubsection*{Point Particles}

We start by expressing the mass coupling, from the worldline point particle in eq. (\ref{eq:actionlab}), in terms of the non-relativistic fields \cite{Kol:2007bc}

\begin{flalign}
   \mathcal{S}_{pp} = -m \int \mathrm{d} t \sqrt{-g_{\mu \nu} \frac{\mathrm{d}x^{\mu}}{\mathrm{d}t} \frac{\mathrm{d}x^{\nu}}{\mathrm{d}t}} = 
     -m \int \mathrm{d}t \left \{ e^{\phi} \sqrt{-(1 - \mathcal{A}_i v^i)^2 + e^{-4\phi} \gamma_{ij} v^{i} v^{j}} \right\}, 
\end{flalign}

\noindent  which is then expanded in the velocity and the KK fields,

\begin{flalign}
\begin{split}
    \mathcal{S}_{pp} = -m \int \mathrm{d}t &\left( 1 - \frac{1}{2}v^2 + \phi  - \mathcal{A}_iv^i - \frac{1}{8}v^4 + \frac{3}{2}\phi v^2 + \frac{1}{2}\phi^2   - \frac{1}{2} v^i v^j \sigma_{ij}  + \;.\;\;.\;\;.\;\;  \right),
\end{split}
\label{eq:sppkk}
\end{flalign}

\noindent where the ellipses denotes higher order corrections. From the last expression we can extract the vertex for the one graviton coupling to the worldline mass,  

\begin{align}
    \begin{split}
        \begin{gathered}
   \includegraphics[width = 1.5cm]{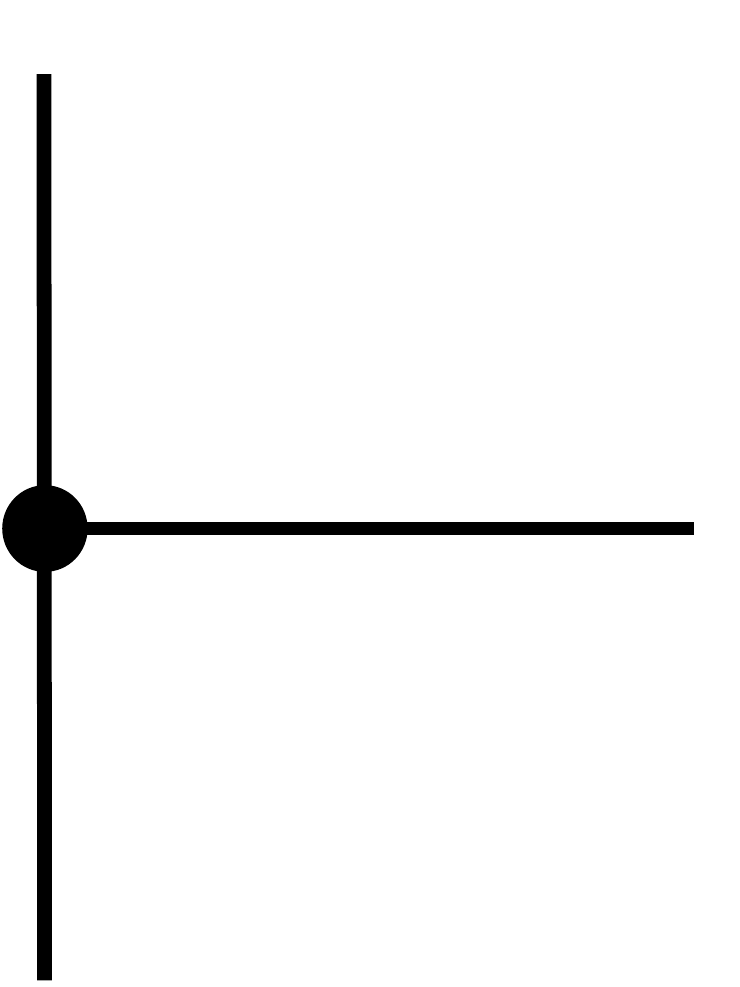}
        \end{gathered}  =& - m \int \mathrm{d} t \phi \left\{ 1 + \frac{3}{2}v^2 + O(v^4) \right\},
    \end{split}
\label{eq:pointmassv}
\end{align}

\noindent where the bold vertical line represents the worldline of the compact object, the black dot represents the mass coupling, and the horizontal line the scalar propagator. 

Higher order corrections can be obtained from the mass coupling with the vector propagator,

\begin{align}
    \begin{gathered}
   \includegraphics[width = 1.5cm]{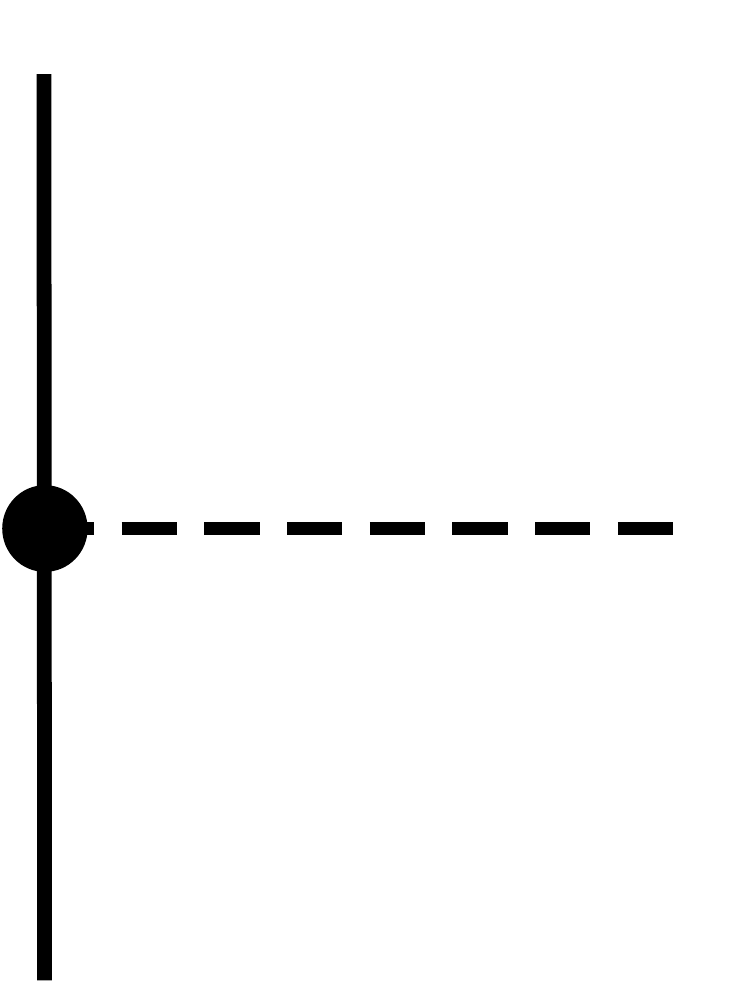}
    \end{gathered}
    =&  \,m \int \mathrm{d}t \mathcal{A}_i v^i \left( 1 + O(v^2) \right).
\end{align}

\noindent The mass coupling with tensor propagator, $\sigma_{ij}$, can be read from eq. (\ref{eq:sppkk}) as well, $\propto v^i v^j \sigma_{ij}$, but is not necessary for the computations of this work. Then, we extract the two graviton exchange,

\begin{align}
    \begin{gathered}
   \includegraphics[width = 1.5cm]{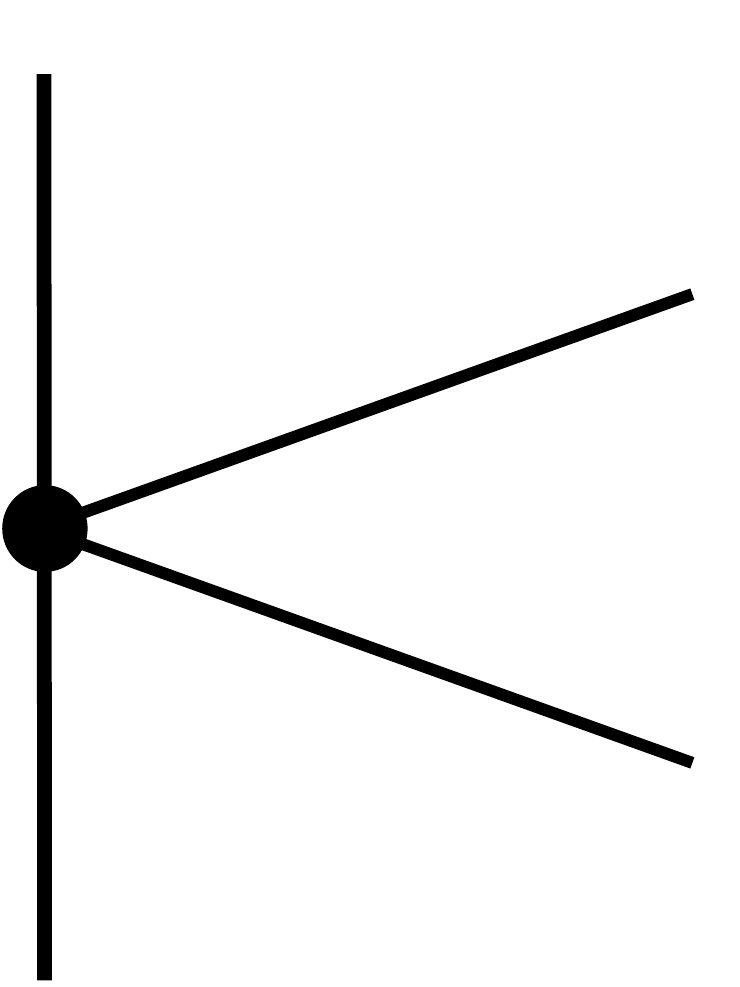}
    \end{gathered}
    ={}& - \frac{1}{2} m \int \mathrm{d}t  \phi^2 \left( 1 + O(v^2) \right).
\end{align}

\noindent Higher order PN corrections to the point mass coupling can be considered from the worldline point particle action \cite{Levi:2018nxp}. 

\subsubsection*{Spinning Particles}

We have cast the minimal spin coupling Lagrangian in a suitable form for the PN expansion in eq. (\ref{eq:lomega2}). In order to extract the correct vertices, it is necessary to take into account the higher order spin correction, $\propto S_{ab} v^a a^b$ \cite{Levi:2015msa}. This term arises by Legendre transforming eq. (\ref{eq:actionlab}), to express it in terms of the spin, such that the relevant corrections reads

\begin{flalign}
\mathcal{L}_{S} =&  \frac{1}{2} S_{ab} \left( \Omega^{ab} + \frac{ a^{a} v^b}{ v^2} \right) = \frac{I}{2} S_{ab} \left( v^{\mu} \omega_{\mu}^{\;ab} + 2(\partial_{\tau} v^a + v^{\mu} \omega_{\mu}^{\;\;ac}v_c)\frac{v^b}{v^2} \right),
\label{eq:lomega4}
\end{flalign}

\noindent  where we have neglected the term, $\propto \Omega \Lambda \partial_{t} \Lambda$, given that it does not contribute to the dynamics.

Before parametrizing the spin Lagrangian in terms of the KK fields, we recall the spin condition of our effective theory \cite{Martinez:2021mkl,Steinhoff:2015ksa},

\begin{flalign}
 \sqrt{v^2} S^{0b} + v_{i} S^{ib} = 0,
\end{flalign}

\noindent which implies the expansion \cite{Levi:2008nh}

\begin{flalign}
S^{0i} = \frac{1}{2} S^{ji} v_j + \frac{1}{4} \Omega^{ji} \mathcal{A}_j + O(v^4),\label{eq:SSCKK}
\end{flalign}

\noindent suppressing the temporal spin components with respect to the spatial ones by at least an order of $v$.

The vertices can be obtained by expressing the spin connection in eq. (\ref{eq:lomega4}), in terms of the vierbein and the Christoffel symbol using eq. (\ref{eq:cristoffele}). Then, by implementing the spin condition, eq. (\ref{eq:SSCKK}), one obtains the expected spin couplings \cite{Levi:2018nxp}

\begin{align}
    \begin{gathered}
   \includegraphics[width = 1.5cm]{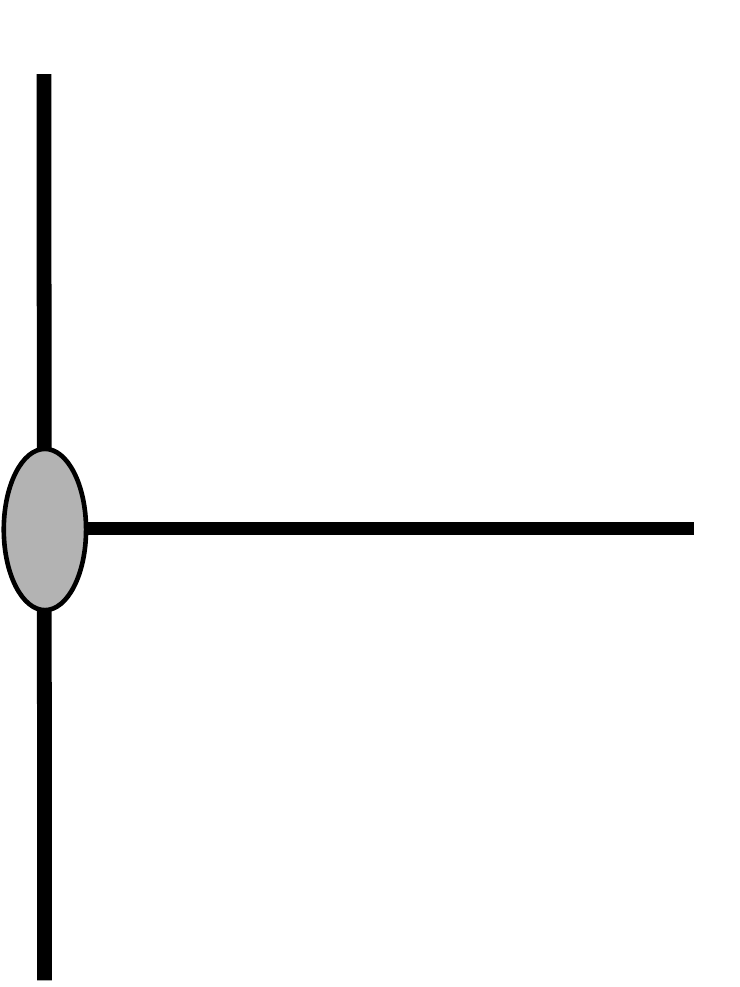}
    \end{gathered}
    =& \int \mathrm{d}t  \epsilon_{ijk} S^k v^i \left(2 \partial^j \phi + v^2 \partial^j \phi + .\;.\;. \right),\label{eq:spinvertexscalar}  
\end{align}

\begin{align}
    \begin{gathered}
   \includegraphics[width = 1.5cm]{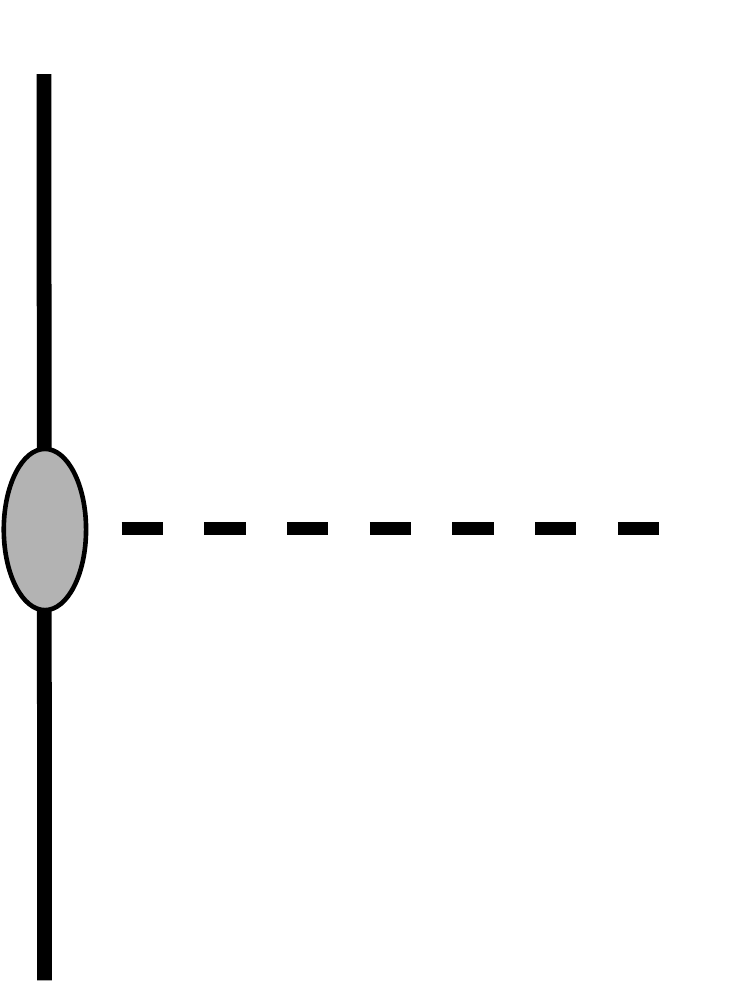}
    \end{gathered}
    =&  \int \mathrm{d}t \epsilon_{ijk} S^k   \left(\frac{1}{2} \partial^i \mathcal{A}^j + \frac{3}{4}v^i v^l (\partial_l \mathcal{A}^j - \mathcal{A}^j \partial_l) + v^i \partial_t \mathcal{A}^j  + .\;.\;. \right), 
\end{align}

\noindent as well as for the spin coupling with tensor propagator, which we do not show here.


\subsubsection*{Size Effects} 

There are two gravitational, electric parity, leading order corrections to consider due to the size of a spinning compact object, one of which is induced by the spin, and the other which is a contribution for any object that has a finite size, even if it is not rotating. The vertices encoding the finite size effects, can be extracted from  

\begin{flalign}
	\mathcal{L}_{W} =   n_{g,\Omega} S^a S^b E_{ab}   + n_{g} E^{ab} E_{ab}.
	\label{eq:Lsize}
\end{flalign}

\noindent  
Therefore, by parametrizing eq. (\ref{eq:Lsize}) in a KK parametrization, we extract the spin-size coupling

\begin{align}
    \begin{gathered}
   \includegraphics[width = 1.5cm]{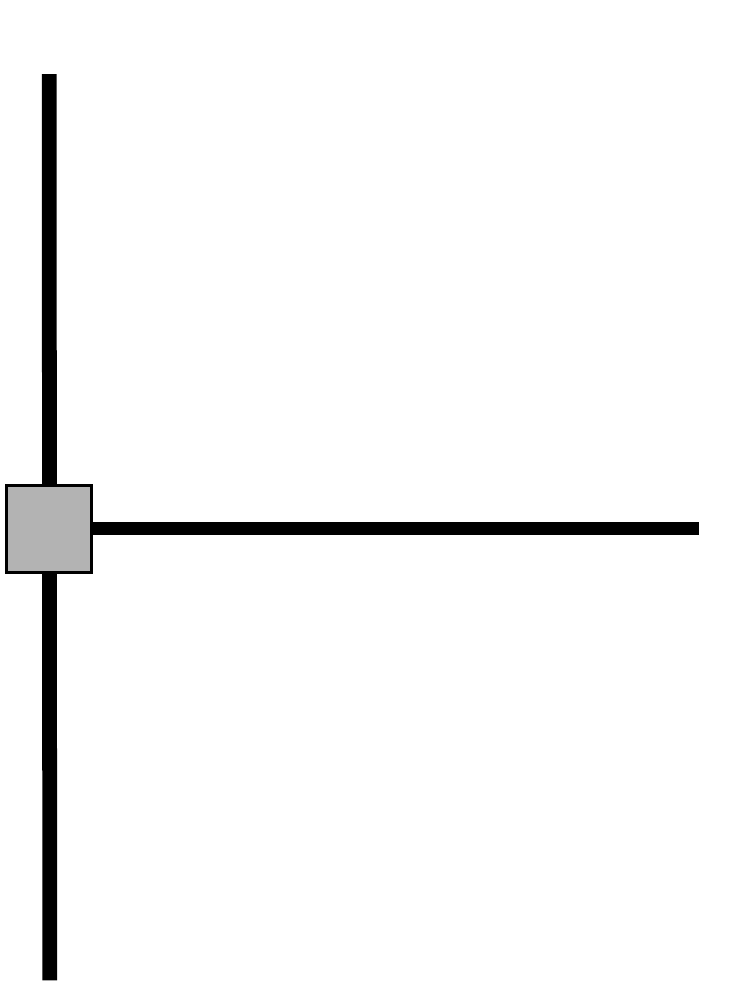}
    \end{gathered}
    =& \int \mathrm{d}t  \frac{n_{g,\Omega}}{2} S^i S^j \partial_i \partial_j \phi +  \;.\;.\;.\,,
\end{align}

\noindent and the two scalar size coupling,

\begin{align}
    \begin{gathered}
   \includegraphics[width = 1.5cm]{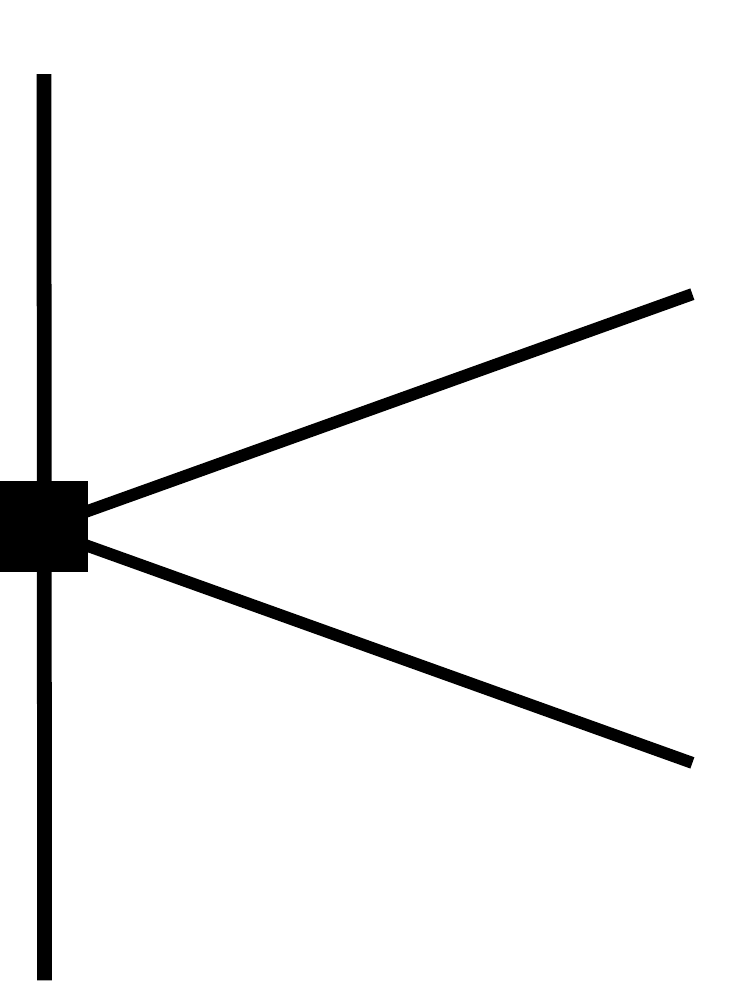}
    \end{gathered}
    ={}&   \int \mathrm{d}t \frac{ n_g}{4} \partial^i \partial^j \phi \partial_i \partial_j \phi + . \; . \; .\,. 
\end{align}

\noindent The coefficients, $n_{g,\Omega}$ and $n_{g}$, encode the internal structure of the object \cite{Martinez:2021mkl}. 

\subsubsection*{Charged Particles}

From the worldline action,

\begin{flalign}
	\mathcal{S}_{q} = \int \mathrm{d} t qv^{a} A_{a} = -\int \mathrm{d} t q A_{0} + \int \mathrm{d}t qv^{i} A_{i},
	\label{eq:chargedpp}
\end{flalign}

\noindent parametrized in the non-relativistic parametrization, the vertices for the photon coupling to the worldline charge can be extracted, 

\begin{align}
    \begin{gathered}
   \includegraphics[width = 1.5cm]{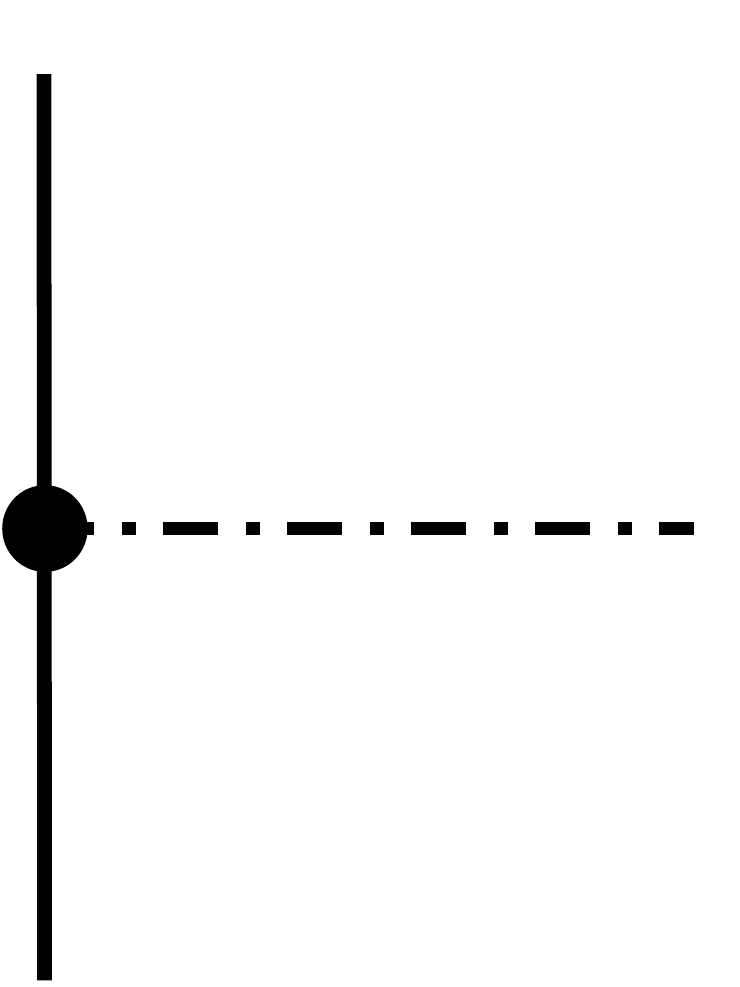}
    \end{gathered}
    = & - q \int \mathrm{d}t  A_0 + .\;.\;.,
\end{align}

\noindent and

\begin{align}
    \begin{gathered}
   \includegraphics[width = 1.5cm]{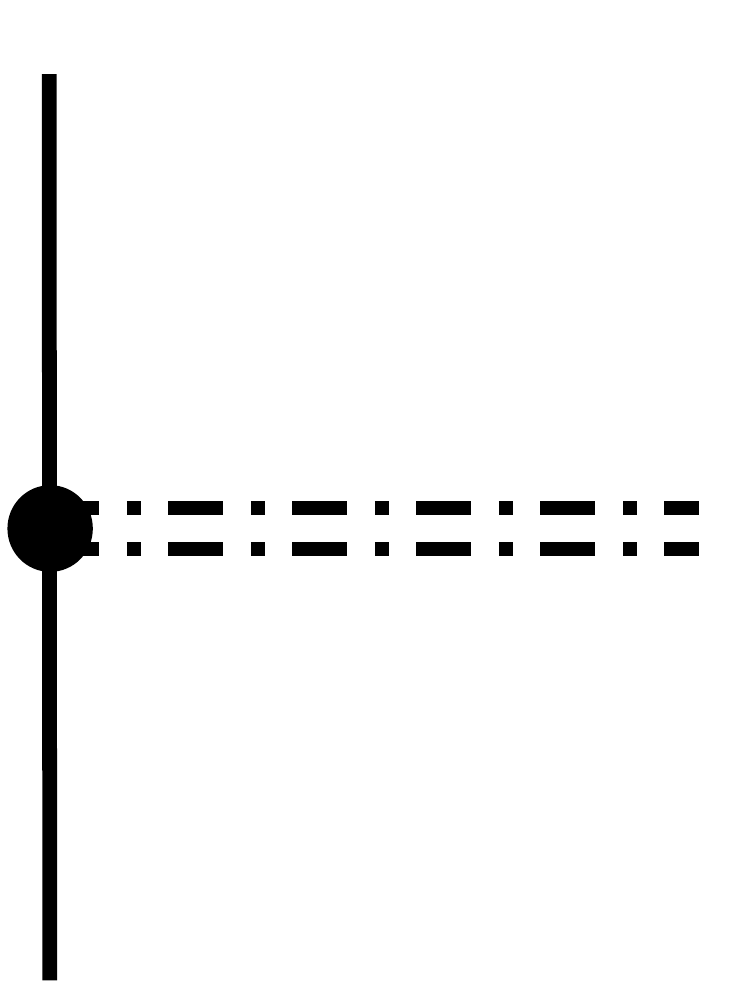}
    \end{gathered}
    ={}&  q \int \mathrm{d}t   A^i v_i + .\;.\;..
\end{align}

\subsubsection*{Polarizability}

An external electromagnetic field in an extended object generates size effects as well, which we will refer as the polarization.  The polarizability is encoded in the terms of the worldline action that are made up of the the electromagnetic tensor, 

\begin{flalign}
	\mathcal{L}_{F} = n_{q,\Omega} S^a B_{a} + n_{q} E^{a} E_{a} + .\,.\,.,
\end{flalign}

\noindent which are invariant under time reversibility. There is a term $\propto B^a B_a$ in the ellipses, but we have neglected to keep the discussion short.  We obtain the scalar spin-polarization coupling 

\begin{align}
    \begin{gathered}
   \includegraphics[width = 1.5cm]{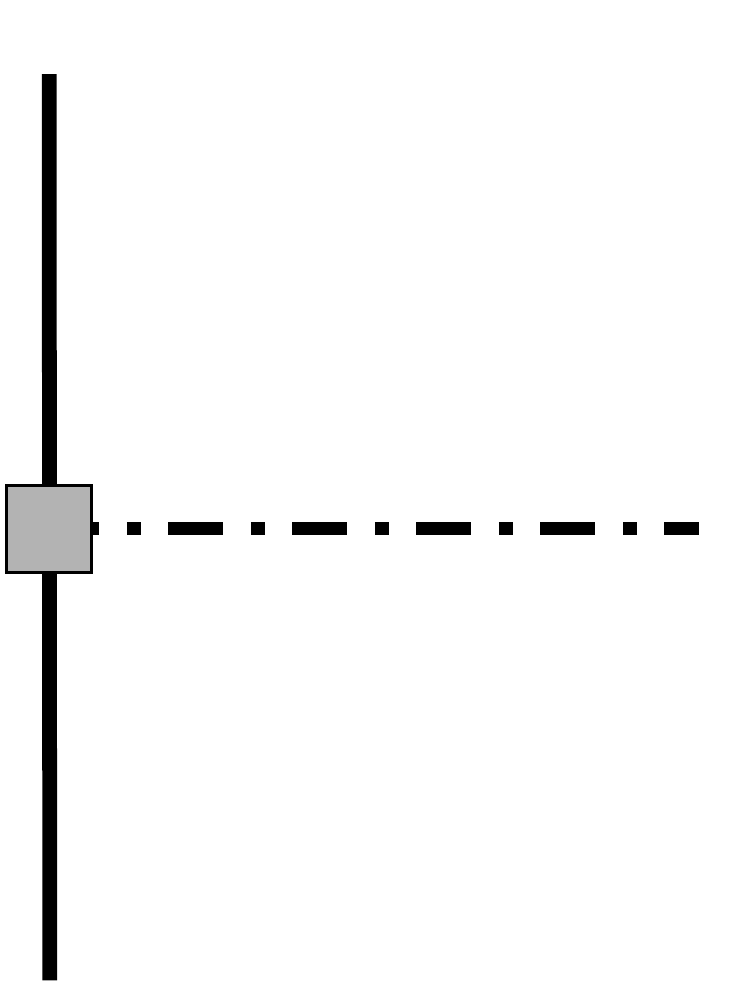}
    \end{gathered}
    = & -  \int \mathrm{d}t  n_{q,\Omega} \epsilon_{ij00} S^i \partial^{
    j} A^0,
\end{align}

\noindent while the two scalar polarization vertex, 

\begin{align}
    \begin{gathered}
   \includegraphics[width = 1.5cm]{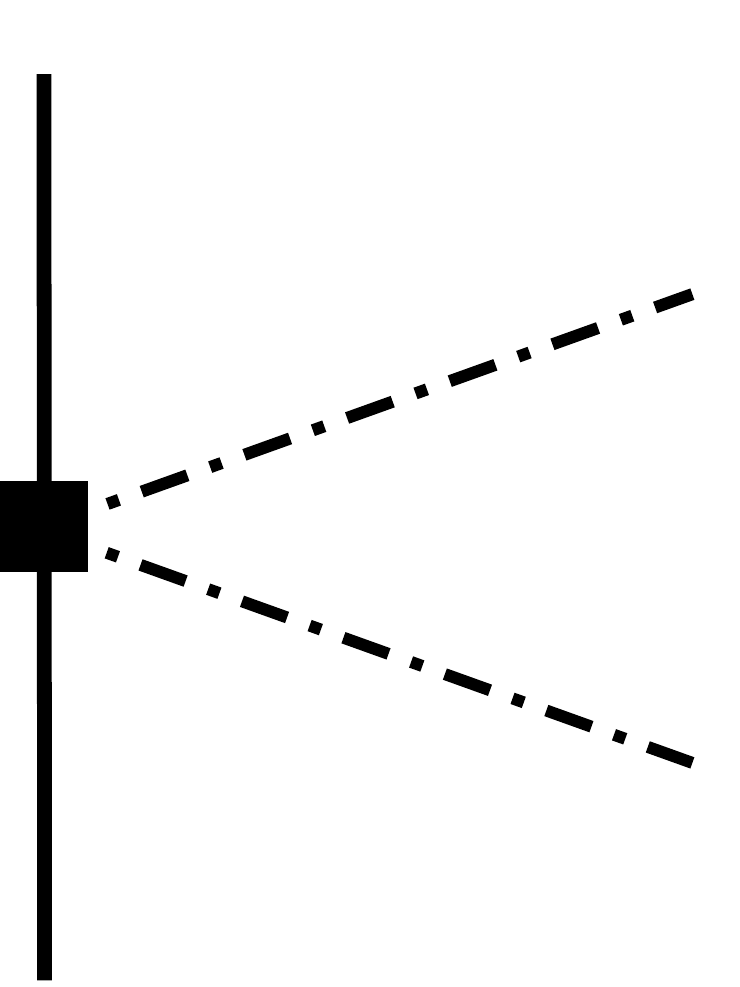}
    \end{gathered}
    ={}&   \int \mathrm{d}t  n_q \partial^i A_0 \partial_i A^0.
\end{align}

\noindent Just as in the gravitational case, the coefficients $n_{q,\Omega}$ and  $n_{q}$, encode the internal structure of the compact object.

\subsubsection*{Dissipation}

The vertices encoding the dissipative effects, can be obtained from the Lagrangian,

\begin{flalign}
    \mathcal{L}_{\mathcal{P},\mathcal{D}} = E_{a}\Lambda^{a}_{\;\; b} \tilde{\mathcal{P}}^b + E_{ab} \Lambda^{a}_{\;\;c} \Lambda^{b}_{\;\;d} \tilde{\mathcal{D}}^{cd} = ic_{q} E_a \frac{\mathrm{d}}{\mathrm{d}t}\Lambda^{a}_{\;\; b}\tilde{E}^{b} + ic_{g} E_{ab} \frac{\mathrm{d} }{\mathrm{d}t}\Lambda^{a}_{\;\;c} \Lambda^{b}_{\;\;d} \tilde{E}^{cd}.
    \label{eq:123}
\end{flalign}

\noindent With the purpose of recovering the LO known results for a slowly rotating object \cite{Endlich:2015mke}, we promote the partial derivative in eq. (\ref{eq:123}) to be covariant, such that 

\begin{flalign}
\frac{\mathrm{D}}{\mathrm{D}t} \Lambda_{c}^{\;a} \Lambda_{d}^{\;b} \tilde{E}^{cd} =&   \left(2 \Lambda_{c}^{\;a} \frac{\mathrm{D}}{\mathrm{D}t} \Lambda_{d}^{\;b} \tilde{E}^{cd} +  \Lambda_{c}^{\;a} \Lambda_{d}^{\;b} \frac{\mathrm{D}}{\mathrm{D}t} \tilde{E}^{ab} \right) \\
= &  \left(2  E^{a}_{\;\;c} \epsilon^{bcd} \Omega_{d}  +   \dot{E}^{ab} + \,.\;\;.\;\;.\, \right).
\end{flalign}

\noindent Therefore, for the gravitational case, there are two LO vertices contributing to the dynamics, the two scalar vertex due to dissipation,

\begin{align}
    \begin{gathered}
   \includegraphics[width = 1.5cm]{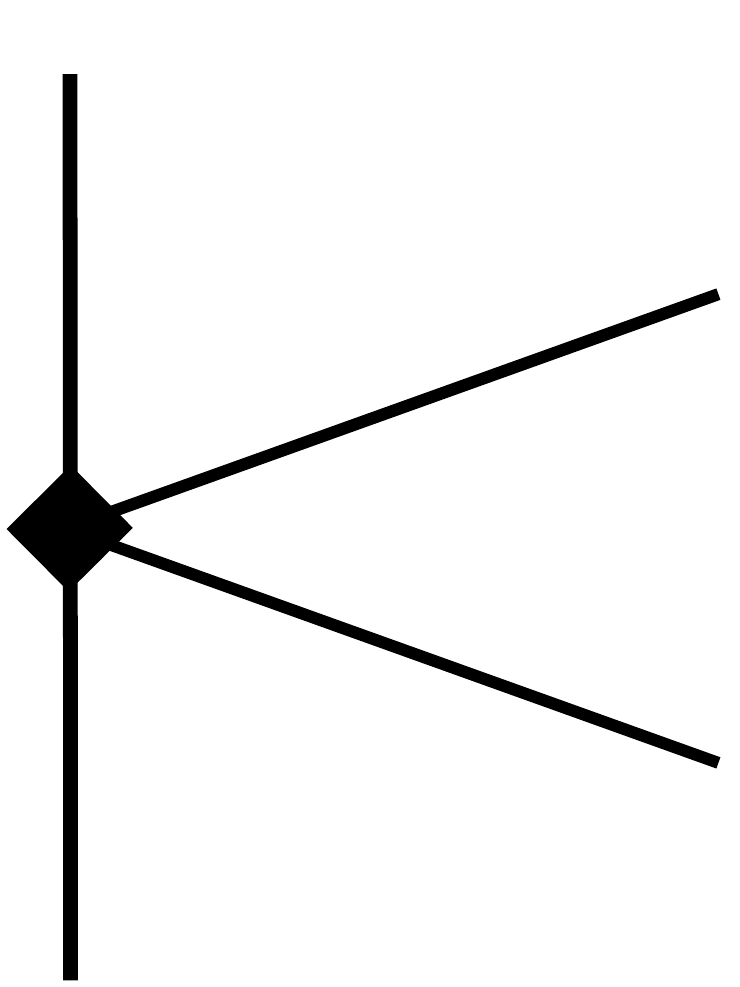}
    \end{gathered}
    ={}&  i \frac{c_g}{4}  \int \mathrm{d}t  \partial_i \partial_j \phi \partial^i \partial^j \dot{\phi}, 
\end{align}

\noindent and the two scalar spin dissipative coupling,

\begin{align}
    \begin{gathered}
   \includegraphics[width = 1.5cm]{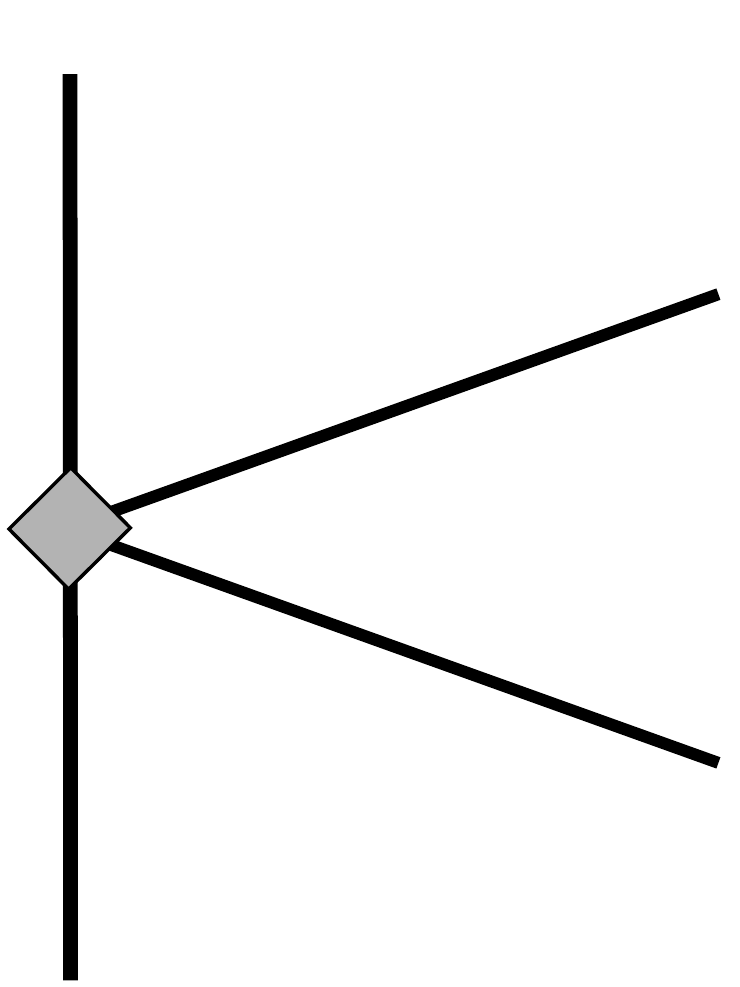}
    \end{gathered}
    ={}&   i \frac{c_{g,\Omega}}{2} \int \mathrm{d}t  \partial_i \partial_j \phi  \partial^{i} \partial_{k} \phi \epsilon^{jkl} S_{l}. 
\end{align}

On the electromagnetic side, a spin dissipative part is not generated in the LO correction. Therefore, the only vertex to be considered, reads

\begin{align}
    \begin{gathered}
   \includegraphics[width = 1.5cm]{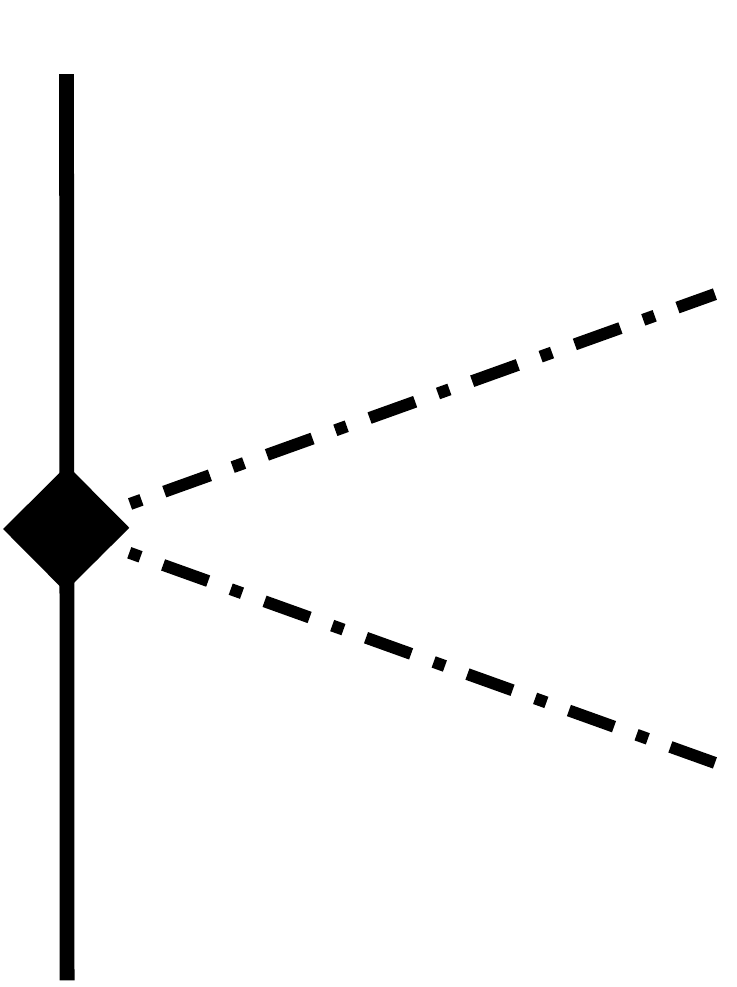}
    \end{gathered}
    ={}&  i c_{q}\int \mathrm{d}t  \partial_i A_0 \partial^i \dot{A}_0.
\end{align}

\subsection{Dynamics of Compact Objects}

In this section we show the computation of the relevant Feynman diagrams to obtain some of the very well known results for the PN expansion, and to derive new LO PN corrections on charged spinning compact objects. Although throughout this work we have worked with units $c = 1$, when showing the final Lagrangian or diagram due to each effect, we will include the $c$, in order to stress the PN order at which the effect enters into the dynamics. Finally, we point out that in the computation of the diagrams, we connect two vertices with the  $\times$ symbol, i.e.

\[ \int \mathrm{d}t_1 \tikzmark{starta} \phi(x_1) \times  \int \mathrm{d}t_2  \tikzmark{enda} \phi (x_2) = 4 \pi G \int \mathrm{d}t_1 \mathrm{d}t_2 \delta(t_1 - t_2)  \int \frac{\mathrm{d}^3 \vec{k}}{(2\pi)^3} \frac{e^{i \vec{k}\cdot \vec{r}}}{\vec{k}^2}, \]
\JoinUp{0.5}{1}{0.5}{1}{a}

\noindent meaning that the relevant propagator is to be inserted. The time relativistic propagator is denoted as $\times|_{t}$.

\subsubsection*{The Newtonian Potential}

\begin{figure}
    \centering
    \includegraphics[width=0.8\textwidth]{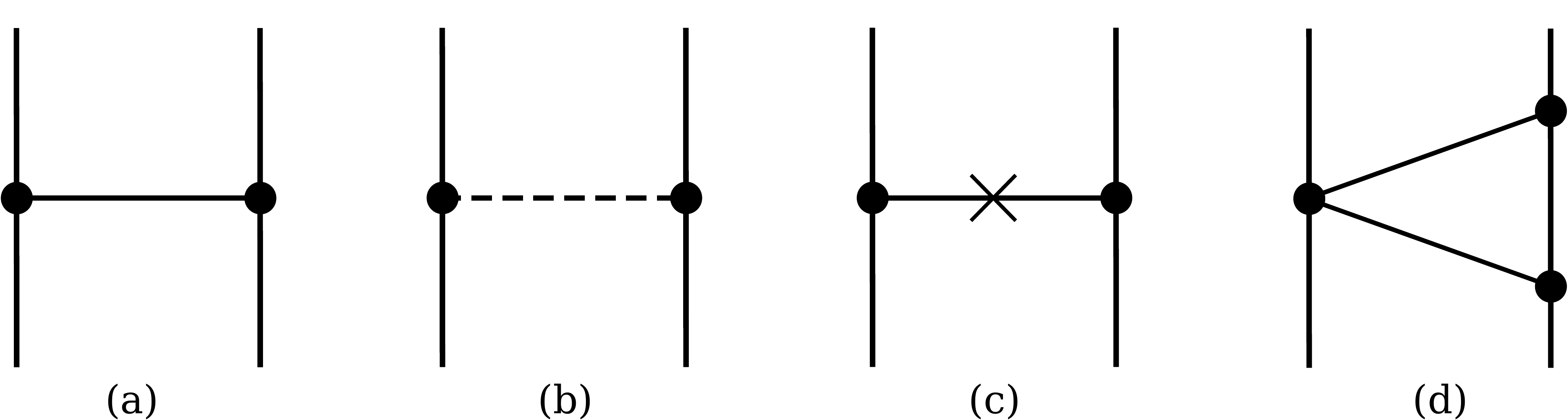}
    \caption{Feynman diagrams to obtain the 0 and 1 PN corrections to the dynamics, from which the Newtonian and the Einstein-Infeld-Hoffman Lagrangian are derived. }
    \label{fig:1PN}
\end{figure}

The simplest example is the Newtonian potential, which is contained in the diagram (a) of figure \ref{fig:1PN}, by neglecting the corrections due to velocity in eq. (\ref{eq:pointmassv}). Therefore, the diagram reads

\begin{flalign}
\begin{split}
	\mathrm{Figure \; } \ref{fig:1PN} \mathrm{(a)} &= m_1 \int \mathrm{d}t_1 \phi(x_1) \times m_2 \int \mathrm{d}t_2 \phi (x_2) \\
    &= 4 \pi G m_1 m_2 \int \mathrm{d}t_1 \mathrm{d}t_2 \delta(t_1 - t_2) \int_{} \frac{\mathrm{d}\vec{k}^3}{(2\pi)^3} \frac{e^{i \vec{k} \cdot \vec{r}}}{\vec{k}^2} = \int \mathrm{d}t \frac{G m_1 m_2}{r},
\end{split}
\label{eq:LN}
\end{flalign}

\noindent with $r = |\vec{x}_1 - \vec{x}_2|$, and where we have used,

\begin{flalign}
	\int_{} \frac{\mathrm{d}\vec{k}^3}{(2\pi)^3} \frac{e^{i \vec{k} \cdot \vec{r}}}{\vec{k}^2} = \frac{1}{4 \pi}\frac{1}{r}.
\end{flalign}

\noindent The Newtonian correction, eq. (\ref{eq:LN}), is a 0 PN correction.

\subsubsection*{The Einstein-Infeld-Hoffman Correction}

The 1 PN correction, known as the Einstein-Infeld-Hoffman correction \cite{EIH}, in the KK parametrization is encoded in the diagrams of figure \ref{fig:1PN}. For the diagram (a), by considering the velocity corrections in eq. (\ref{eq:pointmassv}) up to order $v^2$, the diagram reads

\begin{flalign}
\begin{split}
	\mathrm{Figure \; } \ref{fig:1PN} \mathrm{(a)}  &=  m_1 \int \mathrm{d}t_1 \phi(x_1) \left(1 + \frac{3}{2} v_1^2 \right) \times m_2 \int \mathrm{d}t_2 \phi (x_2) \left(1 + \frac{3}{2} v_2^2 \right)\\
    &= 4 \pi G m_1 m_2 \int \mathrm{d}t_1 \mathrm{d}t_2 \delta(t_1 - t_2) \int_{\vec{k}} \frac{e^{i \vec{k} \cdot \vec{r}}}{\vec{k}^2} \left(1 + \frac{3}{2} (v_1^2 + v_2^2) + \;.\;.\;.\; \right) \\
    &= \int \mathrm{d}t \frac{G m_1 m_2}{r} \left(1 + \frac{3}{2} (v_1^2 + v_2^2)  \right). 
\end{split}
\end{flalign}

\noindent The next diagram with the vector propagator, yields

\begin{flalign}
\begin{split}
	\mathrm{Figure \; } \ref{fig:1PN} \mathrm{(b)}  &=  m_1 \int \mathrm{d}t_1 \mathcal{A}_{i}(x_1)v^{i}_1 \times m_2 \int \mathrm{d}t_2 \mathcal{A}_{j}(x_2)v^{j}_2\\
    &=  -\int \mathrm{d} t \frac{4 G m_1 m_2}{r} \delta_{ij} v_1^i v_2^j   = -\int \mathrm{d} t \frac{4 G m_1 m_2}{r} (\vec{v}_1 \cdot \vec{v}_2). 
\end{split}
\end{flalign} 

\noindent Then, the diagram with the time relativistic correction 

\begin{flalign}
\begin{split}
	\mathrm{Figure \; } \ref{fig:1PN} \mathrm{(c)} &= m_1 \int \mathrm{d}t_1 \phi(x_1) \times|_{t} \, m_2 \int \mathrm{d}t_2 \phi (x_2)\\
	&= 4 \pi G m_1 m_2 \int \mathrm{d}t_1 \int \mathrm{d}t_2  \frac{\mathrm{d}}{\mathrm{d}t_1 \mathrm{d}t_2} \delta(t_1 - t_2) \int \frac{\mathrm{d}\vec{k}^3}{(2\pi)^3} \frac{e^{\vec{k} \cdot \vec{r}}}{(k^2)^2} \\
    &=    \int \mathrm{d} t \frac{1}{2} \frac{G m_1 m_2}{r} \left( \vec{v}_1 \cdot \vec{v}_2 - \frac{(\vec{v}_1 \cdot \vec{r})(\vec{v}_2 \cdot \vec{r})}{r^2} \right), 
\end{split}
\end{flalign}

\noindent where we have used \cite{Porto:2016pyg},

\begin{flalign}
\int  \frac{\mathrm{d}\vec{k}^3}{(2\pi)^3} \frac{k^i k^j}{\vec{k}^4} e^{i \vec{k} \cdot \vec{r}} = \frac{1}{8 \pi r^3} \left(r^2 \delta^{ij} - r^i r^j \right).	
\label{eq:identtime}
\end{flalign}

Finally, the last diagram which is a non-linear static contribution, reads

\begin{flalign}
\begin{split}
	\mathrm{Figure \; } \ref{fig:1PN} \mathrm{(d)} &= - \frac{1}{2} m_1 \int \mathrm{d}t_1 \phi_1^2 \times m^2_2 \int \mathrm{d}t_2 \phi_2^2\\
	&=- \int \mathrm{d}t  \frac{1}{2} \frac{G^2 m_1 m_2^2}{r^2}. 
\end{split}
\end{flalign}

\noindent By obtaining the mirror image of \ref{fig:1PN}(d), and gathering all of our results, we obtain the 1 PN, Einstein-Infeld-Hoffman Lagrangian,

\begin{flalign}
\begin{split}
\mathcal{L}_{EIH} =& \frac{1}{c^2}\Bigg( \frac{1}{8} \sum_{1,2} m v^4 + \frac{3}{2} \frac{G m_1 m_2}{r} \left( v_1^2 + v_2^2  \right) - \frac{7}{2} \frac{ G m_1 m_2}{r} (\vec{v}_1 \cdot \vec{v}_2) \Bigg. \\ 
&\Bigg. \;\;\;\;\;\;\;\; -\frac{1}{2} \frac{G m_1 m_2}{r}  \frac{(\vec{v}_1 \cdot \vec{r})(\vec{v}_2 \cdot \vec{r})}{r^2}  -   \frac{1}{2} \frac{G^2 m_1 m_2 (m_1 + m_2)}{r^2} \; \Bigg), 
\end{split}
\end{flalign}

\noindent where the higher order kinetic term has been obtained from the worldline point particle in eq. (\ref{eq:sppkk}). 

\subsubsection*{Spinning Objects}

\begin{figure}
    \centering
    \includegraphics[width=0.8\textwidth]{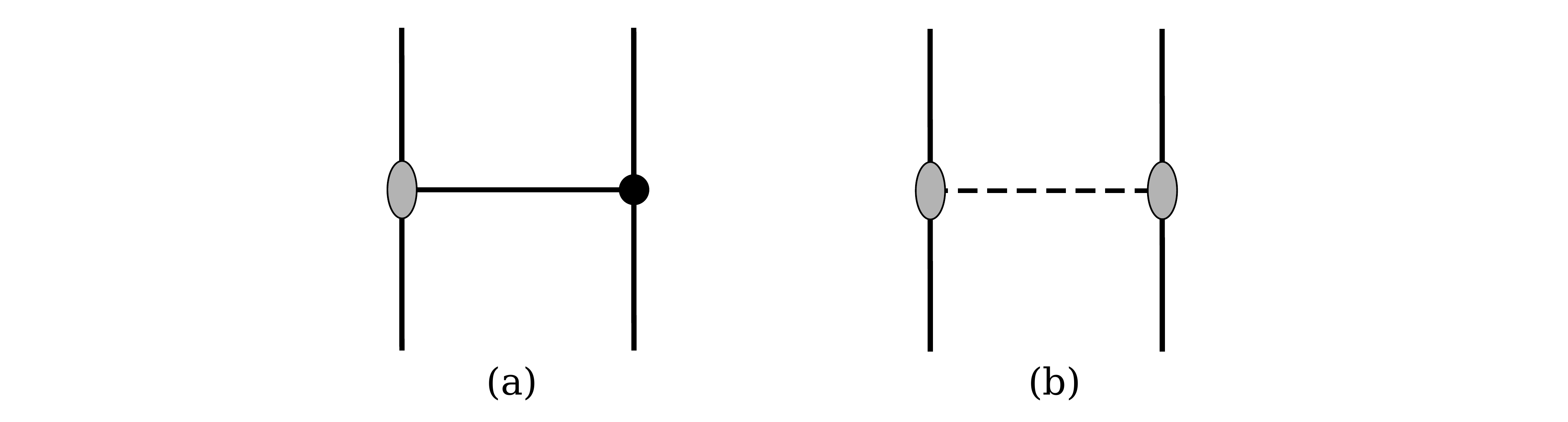}
    \caption{The LO PN correction from the spin-orbit coupling, which is encoded in (a), and the LO PN correction from the spin-spin coupling,  encoded in (b). }
    \label{fig:PNspin}
\end{figure}

For spinning objects, two effects come into play in the dynamics from the spin-orbit and the spin-spin coupling. The LO terms are contained in the diagrams of fig. \ref{fig:PNspin}.
We start by considering fig. \ref{fig:PNspin}(a), which yields the well known LO spin orbit contribution,

\begin{flalign}
\begin{split}
	\mathrm{Figure} \; \ref{fig:PNspin} \mathrm{(a)} =& - \int \mathrm{d}t_1 2 \epsilon_{ijk} S^{k}_1 v^i_1 \partial^j \phi (x_1) \times  \int \mathrm{d}t_2 m_2 \phi (x_2)\\
	=& -  \int \mathrm{d} t G m_2 2 \epsilon_{ijk} S^{k}_1 v^i_1 \partial^j \frac{1}{r}\\
	=&\; \int \mathrm{d}t \frac{2}{c^3} \frac{G m_2}{r^2} \vec{S}_1 \cdot (\vec{v}_{1} \times \hat{r}),
\end{split}
\label{eq:Lso}
\end{flalign}

\noindent with $\hat{r}$, the unit radial vector, and where we have used, $\epsilon_{ijk}a^ib^jc^k = a \cdot (b \times c)$. To obtain the dynamics of the binary, the mirror image of the last diagram is needed as well. 

The next LO correction, from the spin-spin coupling, is encoded in fig. \ref{fig:PNspin}(b). This diagram yields 

\begin{flalign}
\begin{split}
	\mathrm{Figure} \; \ref{fig:PNspin}\mathrm{(b)} =&  \int \mathrm{d}t_1 \frac{1}{2} \epsilon_{ijk} S^{k}_1  \partial^i A^j (x_1)  \times \int \mathrm{d}t_2 \frac{1}{2} \epsilon_{lmn} S^{n}_2 \partial^l A^m (x_2)\\
	=& - \int \mathrm{d}t  G \epsilon_{ijk} S^{k}_1 \delta^{jm} \partial^i \partial^l  \left(\frac{1}{r} \right) \epsilon_{lmn} S^n_2 \\
	=& -\int \mathrm{d} t \frac{1}{c^4} \frac{G}{r^3} \left( \vec{S}_1 \cdot \vec{S}_2 - 3 \left( \vec{S}_1 \cdot \hat{r} \right)\left( \vec{S}_2 \cdot \hat{r} \right) \right),
\end{split}
\label{eq:Lss}
\end{flalign}

\noindent as expected \cite{Levi:2015msa}. Therefore, the LO spin-orbit contribution, eq. (\ref{eq:Lso}), enters at order 1.5 PN, while the LO spin-spin contribution, (\ref{eq:Lss}), enters at 2 PN, for a maximally spinning compact object. If the star is slowly spinning, then these effects will enter into the dynamics at a higher order.

\subsubsection*{Size Effects}

\begin{figure}
    \centering
    \includegraphics[width=0.8\textwidth]{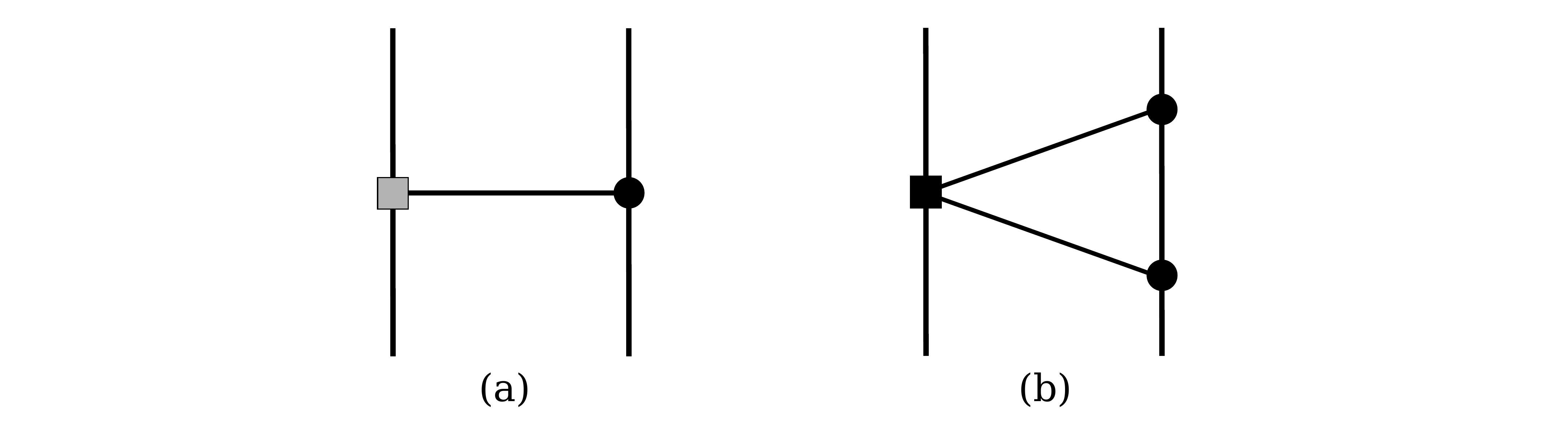}
    \caption{Corrections due to size effects. In (a), the LO spin-size correction, while (b), the LO static size correction.}
    \label{fig:PNsize}
\end{figure}

The LO size effects due to gravity are contained in the diagrams of fig. \ref{fig:PNsize}. The correction from the spin-gravity coupling, reads

\begin{flalign}
\begin{split}
\mathrm{Figure} \; \ref{fig:PNsize}\mathrm{(a)} =& \int \mathrm{d}t_1 \frac{n_{\Omega,g}}{2} S_1^i S_1^j \partial_i \partial_j \phi (x_1) \times \int \mathrm{d} t_2 m_2 \phi (x_2) \\
=& \int \mathrm{d}t \frac{n_{\Omega,g}}{2}\frac{G m_2}{r^3} \left(-\vec{S}^2_1 + 3 \left( \vec{S}_1 \cdot \hat{r} \right)^2 \right).
\end{split}
\label{eq:spintides}
\end{flalign}

\noindent The second diagram encodes the size effects of an extended object, even if it is not rotating. We obtain, 

\begin{flalign}
\begin{split}
\mathrm{Figure} \; \ref{fig:PNsize}\mathrm{(b)} =& \int \mathrm{d}t_1 \frac{n_{g}}{4} \partial^i \partial^j  \phi(x_1) \partial_i \partial_j \phi (x_1) \times \int \mathrm{d} t_2 m_2 \phi (x_2)m_2 \phi (x_2) \\
=& \int \mathrm{d}t  3n_{g}\frac{G^2 m_2^2}{r^6}.
\end{split}
\label{eq:statictides}
\end{flalign}

\noindent Both of the size corrections agree with the literature \cite{Levi:2015msa,Endlich:2015mke}. The mirror image of both diagrams are needed for a complete description of the system.  

For the correction in eq. (\ref{eq:statictides}), due to static tides, it is commonly said that it enters into the dynamics at 5 PN order. This is because the coefficient, $n_g \propto \ell^5/G$, with $\ell$ the radius of the object. For a compact object,  $\ell \sim G m $, for which altogether with the virial theorem, $v^2 \sim Gm/r$,  implies the parameter expansion, $\ell/r \sim v^2$. Therefore, one can identify that eq. (\ref{eq:statictides}), will be of order $v^{10}$, which according to the PN counting, it is equivalent to a 5 PN. Nevertheless, this is not strictly speaking a 5 PN correction, although it helps us to understand the order at which this effect will play a role in the dynamics. 

A similar analysis can be done for eq. (\ref{eq:spintides}). For instance, for a NS, the coefficient $n_{\Omega,g} \sim n_g$ \cite{Yagi:2016bkt}, and therefore one can conclude that the spin-size effects roughly enters at 2 PN order. The magnitude of the spin and the radius of the star will change the order at which this effect enters into the PN expansion.

\subsubsection*{Charge}

\begin{figure}
    \centering
    \includegraphics[width=0.8\textwidth]{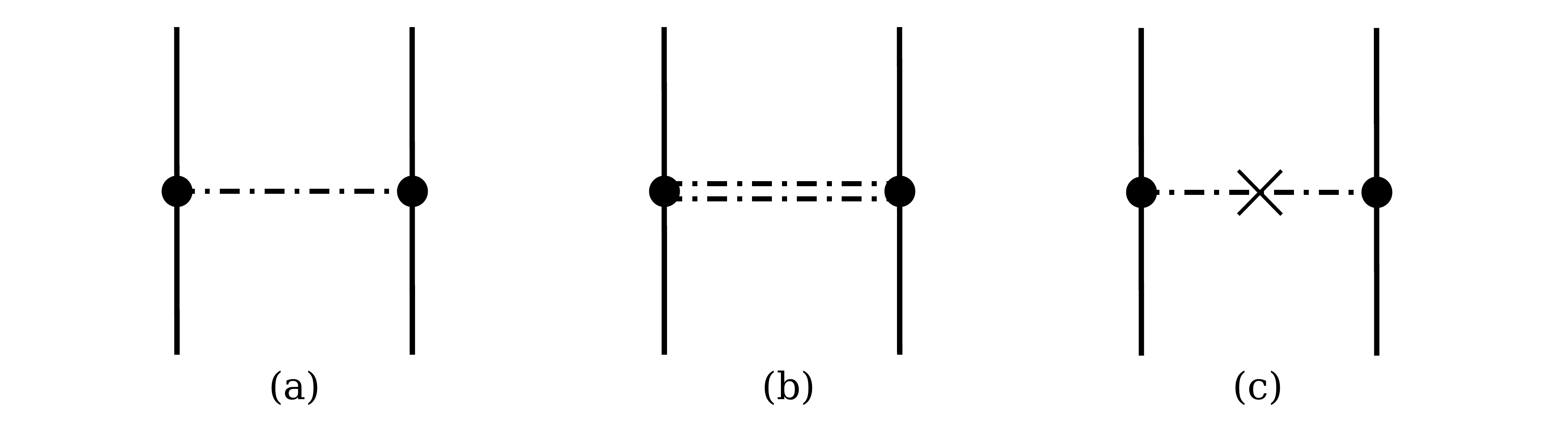}
    \caption{Corrections due to charge. The Coulomb interaction, which is 0 PN order, is encoded in (a), while the 1 PN correction is encoded in the last two diagrams. }
    \label{fig:PNelectro}
\end{figure}

Moving into the dynamics of the electromagnetic interaction, the 0 and 1 PN corrections are encoded in fig. \ref{fig:PNelectro}. We start with the diagram (a), which yields

\begin{flalign}
\begin{split}
\mathrm{Figure} \; \ref{fig:PNelectro} \mathrm{(a)} =& \int \mathrm{d}t_1 q_1 A_0 (x_1) \times \int \mathrm{d}t_2 q_2 A_0 (x_2) \\
=& - \int \mathrm{d}t \frac{\mu_0 }{4 \pi} \frac{q_1 q_2}{r},
\end{split}
\end{flalign}

\noindent the Coulomb potential, which is 0 PN order. 

Then, we obtain the diagram (b),

\begin{flalign}
\begin{split}
\mathrm{Figure} \; \ref{fig:PNelectro} \mathrm{(b)} =& \int \mathrm{d}t_1 q_1 A^i (x_1) v_{1i}\times \int \mathrm{d}t_2 q_2 A^j (x_2) v_{2j}\\
=& \frac{\mu_0}{4 \pi} \frac{q_1 q_2}{ r} \left(\vec{v}_1 \cdot \vec{v}_2 \right). 
\end{split}
\end{flalign}

\noindent The diagram (c), with the time relativistic correction, yields

\begin{flalign}
\begin{split}
	\mathrm{Figure \; } \ref{fig:PNelectro} \mathrm{(c)} &= \int \mathrm{d}t_1 q_1 A_0 (x_1)  \times |_{t} \, \int \mathrm{d}t_2 q_2 A_0 (x_2)  \\
	&= - \mu_0 q_1 q_2 \int \mathrm{d}t_1 \int \mathrm{d}t_2  \frac{\mathrm{d}}{\mathrm{d}t_1 \mathrm{d}t_2} \delta(t_1 - t_2) \int \frac{\mathrm{d}\vec{k}^3}{(2\pi)^3} \frac{e^{\vec{k} \cdot \vec{r}}}{(k^2)^2} \\
    &=   - \int \mathrm{d} t \frac{\mu_0}{2} \frac{q_1 q_2}{4 \pi}  \frac{1}{r} \left( \vec{v}_1 \cdot \vec{v}_2 - \frac{(\vec{v}_1 \cdot \vec{r})(\vec{v}_2 \cdot \vec{r})}{r^2} \right),
\end{split}
\end{flalign}

\noindent where we used eq. (\ref{eq:identtime}).

Gathering our results, we obtain the 1 PN correction of the electromagnetic interaction \cite{Patil:2020dme}

\begin{flalign}
\mathcal{L}_{q,1\mathrm{PN}}  =  \frac{1}{c^2} \left\{    \frac{\mu_0}{2} \frac{q_1 q_2}{4 \pi} \frac{1}{r}\left( \left(\vec{v}_1 \cdot \vec{v}_2 \right) + \frac{(\vec{v}_1 \cdot \vec{r})(\vec{v}_2 \cdot \vec{r})}{r^2} \right) \right\}.
\end{flalign}

\noindent The non-linear interaction between the gravitational and electromagnetic forces that has been obtained in \cite{Patil:2020dme}, can be obtained by considering higher order corrections in the worldline point particle that describes the charged point particle in the non-relativistic parametrization, i.e. $\propto  A_0 \phi$. 

\subsubsection*{Polarizability}

\begin{figure}
    \centering
    \includegraphics[width=0.8\textwidth]{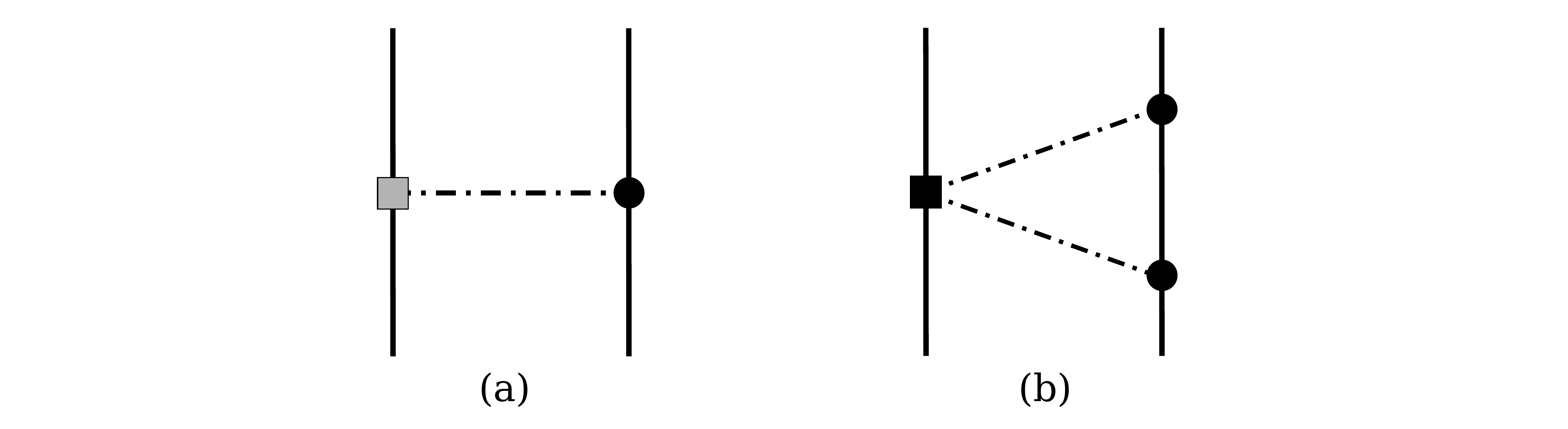}
    \caption{The LO correction due to the polarization for a spinning compact object in diagram (a), and for a non-spinning in diagram (b).}
    \label{fig:PNelectrosize}
\end{figure}

The polarization induced by an electromagnetic field is encoded in the diagrams of fig \ref{fig:PNelectrosize}. For an object that is spinning, the LO magnetic polarization is given by

\begin{flalign}
\begin{split}
\mathrm{Figure} \; \ref{fig:PNelectrosize} \mathrm{(a)} =& \int \mathrm{d}t_1 n_{q_1,\Omega_1} \epsilon_{ij00} S^i_1 \partial^{j} A_0 (x_1) \times \int \mathrm{d}t_2 q_2 A_0 (x_2)\\ 
=&  \int \mathrm{d}t \frac{\mu_0}{4 \pi} \frac{q_2 n_{q_1,\Omega_1}}{r^3} \epsilon_{ij00} S^i_1 r^{j},
\label{eq:Ldissqs}
\end{split}
\end{flalign}

\noindent The LO electric polarization correction when an object is not rotating, reads

\begin{flalign}
\begin{split}
\mathrm{Figure} \; \ref{fig:PNelectrosize}\mathrm{(b)} =& \int \mathrm{d}t_1  n_{q_1}  \partial^i A_0(x_1) \partial_i A_0 (x_1)\times \int \mathrm{d}t_2 q_2^2 A_0 (x_2) A_0 (x_2)\\ 
=&  \int \mathrm{d}t n_{q_1}\left(\frac{\mu_0 q_2}{4 \pi}\right)^2 \frac{1}{r^4},
\label{eq:Ldissq}
\end{split}
\end{flalign}

Both diagrams show that an extended object, even without charge, will be polarized if the companion is charged and the coefficients are different from zero, $n_{q...} \neq 0$.
The coefficients, $n_q$ for a BH are zero \cite{Hui:2021vcv,Charalambous:2021mea}, while for a NS are unknown. By dimensional analysis we find that, $n_{q,\Omega} \propto  n_q \propto \ell^3$, which implies that, eq. (\ref{eq:Ldissqs}) and eq. (\ref{eq:Ldissq}), enter roughly at 1 PN and 3 PN respectively. The magnitude of the charge and the spin will change the order at which these effects enter into the dynamics. 

\subsubsection*{Dissipative Effects}

We consider both electromagnetic and gravitational dissipative effects. For a non-spinning BH, dissipation takes into account for the absorption of electromagnetic and gravitational waves. For a non-spinning NS, we can think of this effect as the energy loss due to its equation of state of the matter during an electromagnetic or gravitational interaction. Dissipation on spinning objects, arises given that the spin of the body has a time dependence between the object and its tidal environment.

\begin{figure}
    \centering
    \includegraphics[width=0.8\textwidth]{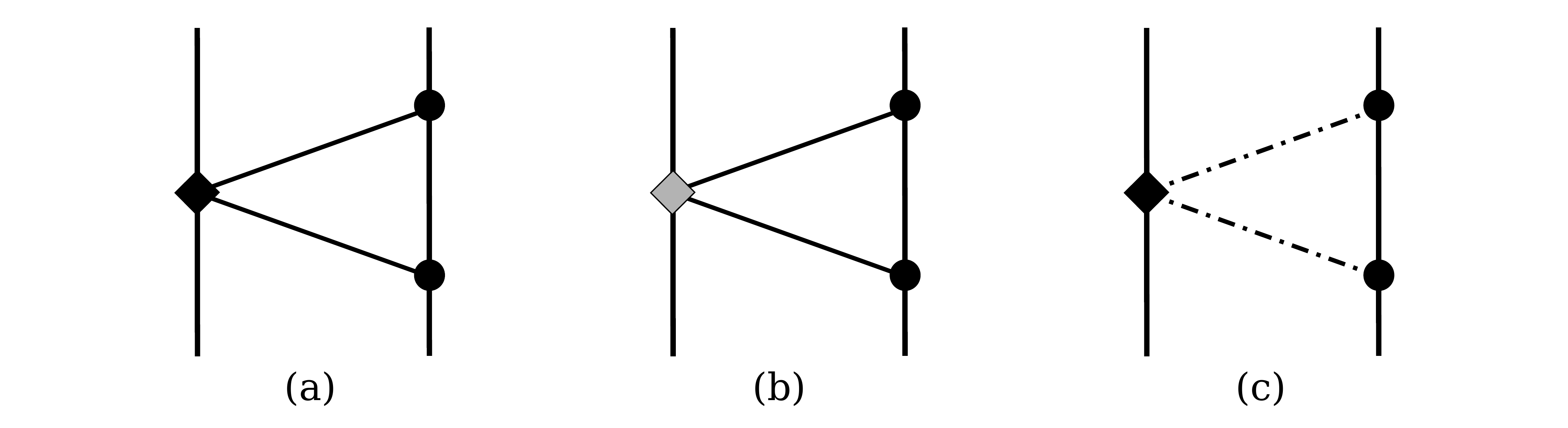}
    \caption{The LO corrections due to dissipative effects. For the gravitational case, the LO correction receives two contributions, (a) and (b), the first one due to the gravitational interaction, and the second one due to spin. For the LO dissipative correction in the electromagnetic case, the spin part does not contribute, thus having only the diagram (c). }
    \label{fig:PNdiss}
\end{figure}

The dissipative correction for the gravitational case, even if the object is not rotating, reads 

\begin{flalign}
\begin{split}
\mathrm{Figure} \; \ref{fig:PNdiss} \mathrm{(a)} =&  \int \mathrm{d}t_1 \frac{ic_{g_1}}{4} \partial_i \partial_j \phi (x_1)  \partial^i \partial^j \dot{\phi}(x_1)   \times  m_2^2 \int \mathrm{d}t_2 \phi^2(x_2)\\
=& \int \mathrm{d}t \frac{i c_{g_1} }{4} \partial_i \partial_j \left(\frac{G m_2}{r} \right) \partial^i \partial^j \partial^t \left(\frac{G m_2}{r} \right)\\
=& \int \mathrm{d}t i 9c_{g_1} G^2 m^2_2 \frac{(\vec{r}\cdot \vec{v})}{r^8} .
\end{split}
\label{eq:dissdiaga}
\end{flalign}

\noindent For a BH, the coefficient, $c_g \propto \ell_{s}^6/G$, due to the absorption of GWs \cite{Goldberger:2005cd}, and therefore its dissipative effects enter at 6.5 PN order.  For a stellar object that is described by an equation of state of matter, in the weak friction model, the coefficient $c_g = \Theta n_g$, with $\Theta$, being the time lag \cite{Hut1981}. Thus, by using the Love coefficient, for which $ n_{g} \propto \ell^5/G$, we conclude that if a NS is described by the weak friction approximation, its dissipative effects will play a role at $\sim 5.5$ PN order.   The intermediate line of eq. (\ref{eq:dissdiaga}), is used to obtain the equations of motion below.

If the object is spinning, the dissipative correction yields
 
\begin{flalign}
\begin{split}
\mathrm{Figure} \; \ref{fig:PNdiss}\; \mathrm{(b)} =&  \int \mathrm{d}t_1 \frac{i c_{g_1,\Omega_1}}{2} \partial_i \partial_j \phi (x_1) \partial^i \partial_k \phi (x_1) \epsilon^{jkl} S_{l}   \times  m_2 \int \mathrm{d}t_2 \phi(x_2) \\
=& \int \mathrm{d} t \frac{i c_{g_1,\Omega_1}}{2} \partial_i \partial_j \left(\frac{G m_2}{r} \right) \partial^i \partial_k \left(\frac{G m_2}{r} \right) \epsilon^{jkl} S_{l} \\
=& \int \mathrm{d} t \frac{i c_{g_1,\Omega_1}}{2} G^2 m_2^2 \left( \frac{\delta_{jk}}{r^6} + 3 \frac{r_j r_k}{r^8} \right) \epsilon^{jkl} S_{l}.\label{eq:dissspin}
\end{split}
\end{flalign}

\noindent For a spinning BH, $n_{g,\Omega} \propto G M^2 \ell_{+}^5/ c^5(\ell_{+} - \ell_{-})$, depends on both of its radi, $\ell_{+}$ and $\ell_{-}$\cite{Chia:2020yla, Martinez:2021mkl}. A careful analysis shows that this effect scales as a $5$ PN order effect, with two prefactors, $\Omega/c$ and $\ell_+/(\ell_+ - \ell_-)$, and where we have considered, $\ell_{+}/r \sim v^2$.  For a NS this coefficient is  unknown.

In what follows, we derive the equations of motion due to the dissipative effects. The equations of motion due to the gravitational dissipative effects were derived in the Newtonian limit of a spinning star \cite{Endlich:2015mke}, and therefore our LO corrections for dissipative effects should reproduce the results as well. 
For a dissipative system, the equations of motion can be obtained from the modified variation \cite{Galley:2012hx}, 

\begin{equation}
    \delta \mathcal{S} + i \int \mathrm{d}t  \mathrm{d}t' \delta J_{ab} (t) \tilde{G}^{ab,cd}_R (t - t') J_{cd}(t') = 0, 
    \label{eq:modified}
\end{equation}

\noindent where $J_{cd} =   \Lambda^{e}_{\;c} \Lambda^f_{\;d}E_{ef}$, and $\tilde{G}_R$, the retarded correlation function of the operators $\tilde{\mathcal{D}}^{ab}$ \cite{Goldberger:2005cd}. 

Then, using eq.  (\ref{eq:dissdiaga}) and (\ref{eq:dissspin}) in the modified variation,  (\ref{eq:modified}), and varying with respect to $x^i$,  one obtains the expected equations of motion due to dissipation for a gravitationally interacting spinning star \cite{Endlich:2015mke},

\begin{flalign}
m_1 \dot{v}_{1i} \Big|_{\mathcal{D}} = - \frac{c_{g_1}}{2}  \partial_{i} \partial_{j} \partial_{k} \Phi_1 \partial^i \partial_j \dot{\Phi}_1 - c_{g_1,\Omega_1}  \partial_{i} \partial_{j} \partial_{k} \Phi_1 \partial^i \partial_k \Phi_1 \epsilon^{jkl} S_{l}
\end{flalign} 

\noindent with $\Phi_1 = G m_2/r$, the Newtonian potential. The mirror image of the last two diagrams is needed for the complete description of the binary.

Moving into the electromagnetic case, the LO dissipative contribution reads

\begin{flalign}
\begin{split}
\mathrm{Figure} \; \ref{fig:PNdiss}\mathrm{(c)} =&  \int \mathrm{d}t_1 i c_{q_1}  \partial_i A_0 (x_1) \partial^i \dot{A}_0 (x_1) \times  \int \mathrm{d}t_2 q_2^2 A_0^2 (x_2)\\
=& \int \mathrm{d}t\, i c_{q_1} \left(\frac{\mu_0 q_2}{4\pi}\right)^2  \partial_i \left(\frac{1}{r} \right) \partial^i \partial^t \left(\frac{1}{r} \right). \\
\end{split}
\end{flalign}

\noindent For a BH, the coefficient due to the absorption of electromagnetic waves, $c_q \propto \ell_{s}^4$ \cite{Goldberger:2005cd,Martinez:2021mkl}, and therefore  enters roughly at 4.5 PN order. Of course this order will change depending on the charge of the object.  The equations of motion from the modified variation leads to, 

\begin{flalign}
m_1 \dot{v}_{1}^i \Big|_{\mathcal{P}} =  c_{q_1} \left( \frac{\mu_0 q_2}{4 \pi} \right)^2 \left(\frac{v^i}{r^6} + \frac{3 (\vec{r}\cdot \vec{v}) r^i}{r^8} \right).
\end{flalign}

\section{Discussion}
\label{sec:discussion}

In this work, we have obtained the LO post-Newtonian expansion of the relevant terms in the effective action that describes the most general compact object that is allowed in a theory of gravity as General Relativity with electrodynamics, which is described by its mass, spin, charge and internal structure.  We have considered the effective theory for spinning extended objects that has been derived using the coset construction \cite{Delacretaz:2014oxa}, as well as the one for charged and spinning \cite{Martinez:2021mkl}. In \cite{Martinez:2021mkl}, the coefficients of the effective theory have been matched from the literature, which allows us to show the predictivity of our theory. 

We have shown that, by matching the coefficients from the currently used effective theories for extended objects \cite{Goldberger:2004jt,Goldberger:2005cd}, which can be spinning \cite{Porto:2005ac,Levi:2015msa} and charged \cite{Patil:2020dme}, our theory reproduces the well known results. In fact, we have shown that the EFT for spinning extended objects in \cite{Delacretaz:2014oxa}, is equivalent to the one in \cite{Levi:2015msa}, when considering higher order corrections of the bosons, from which the spin-acceleration correction arises. Therefore, our work serves as a bridge between the different effective theories, to provide a better understanding of the description of compact objects.

On the electromagnetic interaction, we obtained the Coulomb potential, which is the LO correction of the electromagnetic interaction of charged point particles in curved-spacetime, and obtained the 1 PN correction in the non-relativistic parametrization. Then, by considering spin as well, we have derived new LO results on the finite-size structure of compact  objects interacting electromagnetically, taking into account for the polarization and dissipation of the object. Our derived results are essential to probe whether or not charged compact objects do exists in nature, and this work sets the foundations to model Neutron Stars with strong magnetic fields.

The development of the covariant building blocks of the effective theory for compact objects in \cite{Delacretaz:2014oxa,Martinez:2021mkl}, opens up the possibility to construct the effective theory to all orders, and to obtain higher order PN corrections. Therefore, this work, altogether with \cite{Martinez:2021mkl}, serves on the foundations for obtaining the state of the art PN results for charged spinning compact objects.

\acknowledgments
I.M. is very thankful to J. Steinhoff and R. Patil for very valuable discussions, and to M. Levi and J. Steinhoff for making their code public. We gratefully acknowledge support from the University of Cape Town Vice Chancellor's Future Leaders 2030 Awards program which has generously funded this research and support from  the South African Research Chairs Initiative of the Department of Science and Technology and the NRF.


\bibliography{bib}

\end{document}